\documentclass[9pt,twocolumn,twoside]{pnas-new}
\templatetype{pnasresearcharticle} % Choose template
\usepackage{accents}
\usepackage{mathtools}

\definecolor{Red}{rgb}{0,0,0}
\newcommand{\lina}[1]{\textcolor{Redt}{{#1}}}

\newcommand{\linat}[1]{\textcolor{Red}{{#1}}}

\definecolor{Redt}{rgb}{0,0,0}
\newcommand{\linajk}[1]{\textcolor{Redt}{{#1}}}

\definecolor{Blue}{rgb}{0,0,0}
\newcommand{\linak}[1]{\textcolor{Blue}{{#1}}}

\newcommand{\Imag}{{\Im\mathrm{m}}}   % Imaginary part 
\newcommand{\Real}{{\mathrm{Re}}}   % Real part

\newcommand{\ve}{\boldsymbol} % Vector
\newcommand{\vek}{\boldsymbol{k}} % Vector k

\DeclareMathSymbol{\widetildesym}{\mathord}{largesymbols}{"65}
\newcommand\lowerwidetildesym{%
	\text{\smash{\raisebox{-1.7ex}{%
				$\widetildesym$}}}}

\newcommand\parwidetilde[1]{%
	\mathchoice
	{\accentset{\displaystyle\scalebox{.3}{(}\lowerwidetildesym\scalebox{.3}{)}}{#1}}
	{\accentset{\textstyle\scalebox{.3}{(}\lowerwidetildesym\scalebox{.3}{)}}{#1}}
	{\accentset{\scriptstyle\scalebox{.3}{(}\lowerwidetildesym\scalebox{.3}{)}}{#1}}
	{\accentset{\scriptscriptstyle\scalebox{.3}{(}\lowerwidetildesym\scalebox{.3}{)}}{#1}}
}

\newcommand\varpm{\mathbin{\vcenter{\hbox{%
				\oalign{\hfil$\scriptscriptstyle+$\hfil\cr
					\noalign{\kern-0.5ex}
					$\scriptscriptstyle\scalebox{.5}{(}{\scriptscriptstyle-}\scriptscriptstyle\scalebox{.5}{)}$\cr}%
}}}}
\newcommand\varmp{\mathbin{\vcenter{\hbox{%
				\oalign{\hfil$\scriptscriptstyle-$\hfil\cr
					\noalign{\kern-0.5ex}
					$\scriptscriptstyle\scalebox{.5}{(}{\scriptscriptstyle+}\scriptscriptstyle\scalebox{.5}{)}$\cr}%
}}}}

\newcommand\horpm{\mathbin{\vcenter{\hbox{$\scriptscriptstyle+$$\scriptscriptstyle\scalebox{.5}{(}$$\scriptscriptstyle-$$\scriptscriptstyle\scalebox{.5}{)}$}}}}%
\newcommand\hormp{\mathbin{\vcenter{\hbox{$\scriptscriptstyle-$$\scriptscriptstyle\scalebox{.5}{(}$$\scriptscriptstyle+$$\scriptscriptstyle\scalebox{.5}{)}$}}}}%

\newcommand\horgl{\mathbin{\vcenter{\hbox{$>$$\scalebox{1}{(}$$<$$\scalebox{1}{)}$}}}}%
\begin{document}

	\title{Optical conductivity of the Majorana mode at the $s$- and $d$-wave topological superconductor edge}
	
	% Use letters for affiliations, numbers to show equal authorship (if applicable) and to indicate the corresponding author
	\author[a,b,c,1]{Lina Johnsen Kamra}
	\author[d]{Bo Lu}
	\author[a]{Jacob Linder}
	\author[e,f]{Yukio Tanaka}
	\author[g,h,1]{Naoto Nagaosa}
	
	\affil[a]{Center for Quantum Spintronics, Department of Physics, Norwegian University of Science and Technology, NO-7491 Trondheim, Norway}
	\affil[b]{Condensed Matter Physics Center (IFIMAC) and Departamento de F\'{i}sica Te\'{o}rica de la Materia Condensada, Universidad Aut\'{o}noma de Madrid, E-28049 Madrid, Spain}
	\affil[c]{Department of Physics, Massachusetts Institute of Technology, Cambridge, MA 02139, USA}
	\affil[d]{Center for Joint Quantum Studies, Tianjin Key Laboratory of Low Dimensional Materials Physics and Preparing Technology,
		Department of Physics, Tianjin University, Tianjin 300354, China}
	\affil[e]{Department of Applied Physics, Nagoya University, Nagoya 464-8603, Japan}
	\affil[f]{Research Center for Crystalline Materials Engineering, Nagoya University, Nagoya 464-8603, Japan}
	\affil[g]{RIKEN Center for Emergent Matter Science (CEMS), Wako, Saitama 351-0198, Japan}
	\affil[h]{Fundamental Quantum Science Program, TRIP Headquarters, RIKEN, Wako 351-0198, Japan}
	
	% Please give the surname of the lead author for the running footer
	\leadauthor{Kamra}
	
	% Please add a significance statement to explain the relevance of your work
	\significancestatement{
		%Authors must submit a 120-word maximum statement about the significance of their research paper written at a level understandable to an undergraduate educated scientist outside their field of speciality. The primary goal of the significance statement is to explain the relevance of the work in broad context to a broad readership. The significance statement appears in the paper itself and is required for all research papers.
		Recent years have seen considerable progress towards realizing non-abelian particles, fueled by their promised applications to topological quantum devices. A prominent example is the one-dimensional chiral Majorana mode. \lina{It opens} the possibility of using wave packets propagating at high speed as an alternative to the braiding of zero-dimensional Majorana fermions. While signatures of the latter have been established, a weak spot in detecting chiral Majorana modes lies in reliably capturing quantitative measures such as a quantized conductivity. \lina{We here propose using microwave spectroscopy to instead reveal distinct \textit{qualitative} signatures emerging due to the unique dispersion of the Majorana mode that allows photons to break up Cooper pairs into Majorana fermions propagating along a topological superconductor edge.}}
	% Please include corresponding author, author contribution and author declaration information
	\authorcontributions{N.N. and Y.T. designed the research. L.J.K. performed the research.  L.J.K. and N.N. wrote the paper. All authors contributed to the understanding and interpretation of the results, and in revising the paper.}
	\correspondingauthor{\textsuperscript{1}To whom correspondence should be addressed. E-mail: ljkamra@mit.edu, nagaosa@riken.jp}
	
	% At least three keywords are required at submission. Please provide three to five keywords, separated by the pipe symbol.
	\keywords{Chiral Majorana edge mode $|$ Topological superconductors $|$ Local optical conductivity}
	
	\begin{abstract}
		%Please provide an abstract of no more than 250 words in a single paragraph. Abstracts should explain to the general reader the major contributions of the article. References in the abstract must be cited in full within the abstract itself and cited in the text.
		The Majorana fermion offers fascinating possibilities such as non-Abelian statistics and non-local robust qubits, and hunting it is one of the most important topics in current condensed matter physics. Most of the efforts have been focused on the Majorana bound state at zero energy in terms of scanning tunneling spectroscopy searching for the quantized conductance. On the other hand, a chiral Majorana edge channel appears at the surface of a three-dimensional topological insulator when engineering an interface between proximity-induced superconductivity and ferromagnetism. Recent advances in microwave spectroscopy of topological edge states open a new avenue for observing signatures of such Majorana edge states through the local optical conductivity. As a guide to future experiments, we show how the local optical conductivity and density of states present distinct qualitative features depending on the symmetry of the superconductivity, that can be tuned via the magnetization and temperature. In particular, the presence of the Majorana edge state leads to a characteristic non-monotonic temperature dependence achieved by tuning the magnetization.
	\end{abstract}
	
	\dates{This manuscript was compiled on \today}
	\doi{\url{www.pnas.org/cgi/doi/10.1073/pnas.XXXXXXXXXX}}

	\maketitle
	\ifthenelse{\boolean{shortarticle}}{\ifthenelse{\boolean{singlecolumn}}{\abscontentformatted}{\abscontent}}{}
	
	\firstpage[3]{4}
	% Use \firstpage to indicate which paragraph and line will start the second page and subsequent formatting. In this example, there are a total of 11 paragraphs on the first page, counting the first level heading as a paragraph. The value {12} represents the number of the paragraph starting the second page. If a paragraph runs over onto the second page, include a bracket with the paragraph line number starting the second page, followed by the paragraph number in curly brackets, e.g. "\firstpage[4]{11}".

	% If your first paragraph (i.e. with the \dropcap) contains a list environment (quote, quotation, theorem, definition, enumerate, itemize...), the line after the list may have some extra indentation. If this is the case, add \parshape=0 to the end of the list environment.
	
	\dropcap{S}tudy of Majorana fermions in topological superconductors is an important topic from the viewpoints of fundamental physics and applications to quantum computing  \cite{Alicea_RepProgPhys_2012,Flensberg_NatRevMater_2021,Yazdani_Science_2023,Tanaka_JPhysSocJpn_2011}. Zero energy Majorana bound states \cite{Kitaev_PhysUspekhi,Fu_PRL_2008,Tanaka_ProgTheorExpPhys} are proposed to act as topologically protected qubits with non-Abelian statistics \cite{Nayak_RevModPhys_2008,Sato_JPhysSocJpn_2016}. While significant progress has been made in experimentally establishing their signatures \cite{Mourik_Science_2012,Rokhinson_NatPhys_2012,Das_NatPhys_2012,NadjPerge_Science_2014,Sun_PRL_2016,Wang_Science_2018,Jack_Science_2019,Manna_PNAS_2020}, another possible Majorana fermion -- the propagating one at the edge of the sample in topological superconductors -- has remained more elusive \cite{PhysRevB.64.054514,Kayyalha_Science_2020,Jack_NatRevPhys_2021}. 
	The one-dimensional nature of these chiral Majorana channels introduce the possibility of using wave packets propagating at high speed as an alternative to the braiding of zero-dimensional Majorana particles \cite{Lian_PNAS_2018}. One predicted platform is the heterojunction of a ferromagnetic insulator and superconductor on top of a three-dimensional (3D) topological insulator \cite{Fu_PRL_2009,Akhmerov_PRL_2009,Tanaka_PRL_2009}. The proximity effect of the magnetization and superconductivity to the surface state of the topological insulator results in the chiral Majorana edge channel at the interface \cite{Fu_PRL_2009,Akhmerov_PRL_2009,Tanaka_PRL_2009,Lu_SupercondSciTech_2015,He_CommunPhys_2019}. Even though the Majorana bound state is charge-neutral, it is coupled to the electromagnetic field due to being a composite quasiparticle rather than a fundamental one, contributing to the optical conductivity \cite{He_PRL_2021,He_PRB_2021,Lu_PRB_2022,Bi_arXiv_2023}. This is particularly promising because the experimental advances in the local probe of microwave spectroscopy \cite{Barber_NatRevPhys_2022,doi:10.1126/sciadv.abd1919} have enabled the detection and imaging of the edge channels in the quantum anomalous Hall \cite{Allen_PNAS_2019,Wang_PRB_2023} and Weyl state \cite{Ma_Science_2015}. 
	
	In the present paper, we study theoretically the local density of states (LDOS) and local optical conductivity near the interface \lina{between proximity-induced ferromagnetism and $s$-wave or $d$-wave superconductivity at the surface of a 3D topological insulator.} They show a variety of behavior as a function of energy depending on the strength of the magnetic exchange coupling and temperature. Especially the contribution of the nodes in the $d$-wave superconductors makes the behavior qualitatively distinct from the $s$-wave case with a full gap. While previous scanning tunneling microscopy (STM) experiments relied on quantitative measures, \textit{e.g.} a quantized conductance \cite{Chung_PRB_2011,Wang_PRB_2015,Lian_PRB_2016}, that cannot be captured exactly due to measurement noise \cite{Ji_PRL_2018,Huang_PRB_2018,Kayyalha_Science_2020,Jack_NatRevPhys_2021}, our results instead provide more robust qualitative features which depend on tuning parameters that can be probed through tunneling spectroscopy and microwave impedance microscopy (MIM) \cite{Allen_PNAS_2019}.
	
	\section*{Results}
	
	\subsection*{Model}
	
\begin{figure}[b!]
		\centering
		\includegraphics[width=\columnwidth]{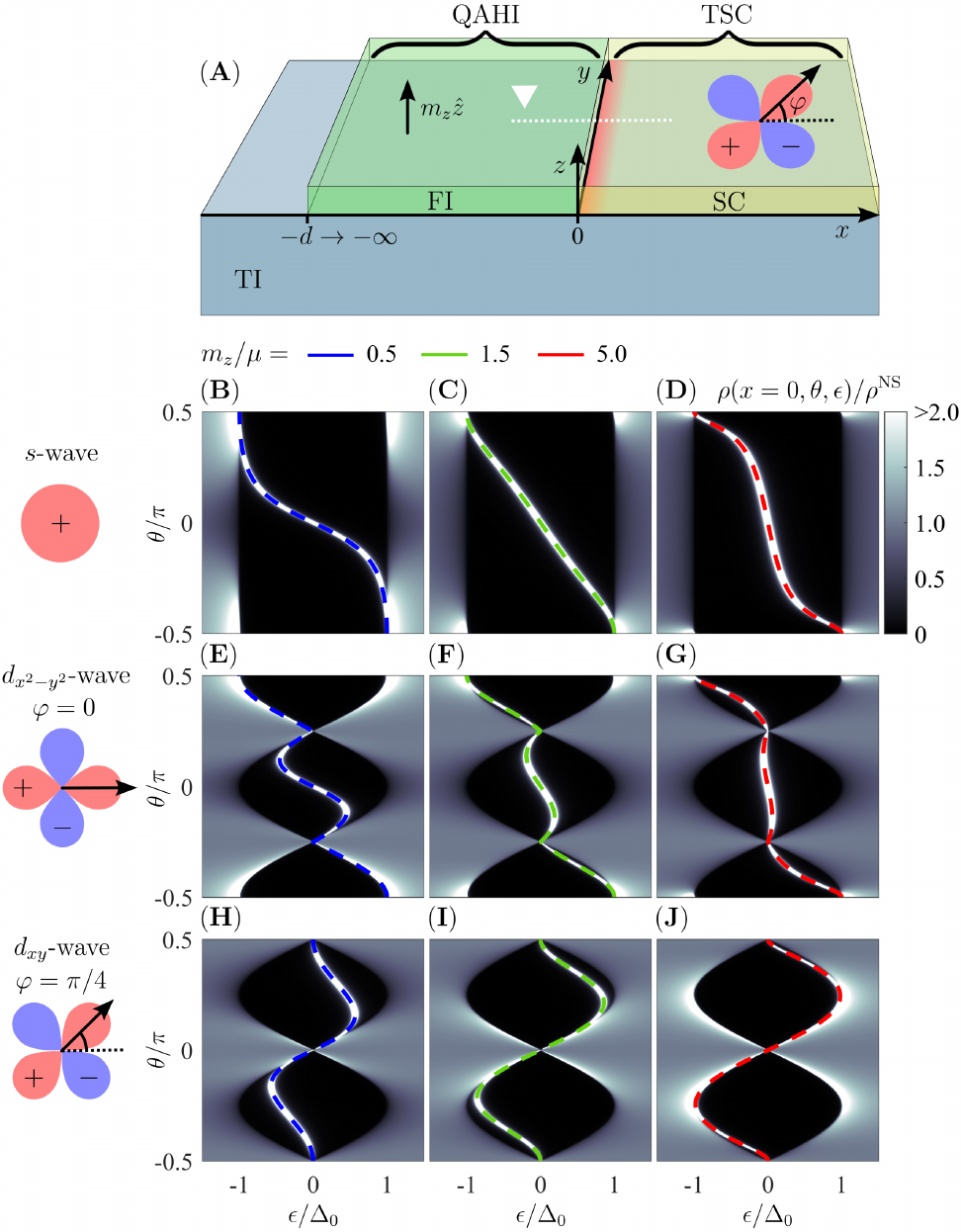}
		\caption{\textbf{Edge states at a QAHI/TSC interface.} (\textbf{A}) The interface between a quantum anomalous Hall insulator (QAHI) and a topological superconductor (TSC) can be studied by proximitizing \lina{the surface of a three-dimensional} topological insulator (TI) to a ferromagnetic insulator (FI) and a superconductor (SC), respectively. At the interface between the 2D QAHI and 2D TSC, there is a single chiral Majorana edge channel (red) decaying over the superconducting coherence length. \lina{The optical conductivity of the edge channel can be measured by running a microwave impedance microscopy (MIM) tip (white \linak{triangle}) across the QAHI/TSC interface from above following the white dotted line \cite{Wang_PRB_2023}. We consider a FI with magnetization $m_z$ along $\hat{z}$, and} a TSC with $s$-wave or $d$-wave pairing, where the positive $d$-wave lobe makes an angle $\varphi$ with respect to the interface normal $\hat{x}$. Panel (\textbf{B})-(\textbf{J}) presents the angle resolved LDOS $\rho(x=0,\theta,\epsilon)$ normalized by its normal-state value for $\Delta_0 = 0$ at the QAHI/TSC interface ($x=0$) in the case of $s$-wave [panel (\textbf{B})-(\textbf{D})], $d_{x^2-y^2}$-wave [panel (\textbf{E})-(\textbf{G})], and $d_{xy}$-wave [panel (\textbf{H})-(\textbf{J})] pairing. \lina{We consider $\Delta_0/\mu=10^{-3}\ll1$, energies $\epsilon+i\delta$, where $\delta/\Delta_0=5\cdot10^{-3}$, and different values of $m_z/\mu$, \textit{i.e.} from left to right, $m_z/\mu = 0.5$, $1.5$, and $5.0$}. The bound state energy dispersion $E_{\text{b}}(\theta)$ [from \eqref{eq:bound_state_energy}] is represented by the dotted lines.}
		\label{fig:01}
	\end{figure}

	We consider the surface of a 3D topological insulator (TI) with spin-orbit parameter $A$ and chemical potential $\mu\geq0$ lying in the $xy$ plane. The TI surface can be separated into three regions: For $x>0$, the TI \lina{surface} is in contact with a superconductor (SC) and hosts proximity-induced superconductivity \cite{Wang_Science_2012,Maier_PRL_2012,Williams_PRL_2012,Veldhorst_NatMater_2012,Yang_PRB_2012,Zareapour_NatCommun_2012} with an order parameter \cite{Tanaka_PRL_1995}
	\begin{align}
		\Delta(\theta) &=\begin{cases}
			\Delta_0 \text{ for $s$-wave pairing, and}\\
			\Delta_0\cos\{2[\theta-\varphi]\} \text{ for $d$-wave pairing.}
		\end{cases}\label{eq:pairing_def}
	\end{align}
	\lina{Above, $\Delta_0$ is the maximum value of the superconducting gap, $\theta$ is given by $k_y = k_{\text{F}}\sin(\theta)$, where $k_{\text{F}}$ is the Fermi momentum, and $\varphi$ is the angle of the positive $d$-wave lobe with respect to the interface normal $\hat{x}$ [See Fig. 1(A)].} In particular, $\varphi=0$ corresponds to $d_{x^2-y^2}$-pairing, and $\varphi=\pi/4$ corresponds to $d_{xy}$-wave pairing. For $-d<x<0$, the TI \lina{surface} has an out-of-plane magnetization $m_z$ induced by an adjacent ferromagnetic insulator (FI). \lina{The finite $m_z$ opens an insulating gap in the surface states of the TI. In this region only, we consider $\mu=0$, yielding a purely decaying wave function. Since the McMillan approach \cite{McMillan_PRL_1968} that we will use in the following relies on constructing the Green's function from propagating wave functions incoming from the left and right, we include a region at $x<-d$ where the TI surface is neither superconducting nor magnetic ($\Delta_0 = 0$, $m_z = 0$). This region introduces propagating wave functions incoming from the left that are helpful for correctly constructing the Green's function \cite{Lu_PTA_2018}.} However, we focus on the interface between the ferromagnetic and superconducting regions by assuming the length of the ferromagnetic region to be much larger than the decay length of the wave function inside this region ($d\to\infty)$. As schematically illustrated in Fig.~\ref{fig:01}(\textbf{A}), the \lina{region where the TI surface is interfaced with a FI} forms a 2D quantum anomalous Hall insulator (QAHI), while the \lina{region where it is interfaced with a SC} forms a 2D topological superconductor (TSC) \cite{Qi_PRB_2010,Tokura_NatRevPhys_2019}. At the QAHI/TSC interface ($x=0$), there is a single chiral Majorana edge channel that will be the focus of this work. 
	
	The superconducting and magnetic regions \linajk{of the topological insulator surface} can be described by the Hamiltonians $H_{\text{TI}}+H_{\text{SC}}$ and $H_{\text{TI}}+H_{\text{FI}}$, respectively \cite{Fu_PRL_2008,Burset_PRB_2015}, where
	\begin{align}
		H_{\text{TI}} = & -\mu\sum_{\vek,\sigma}\psi^{\dagger}_{\sigma} (\vek) \psi_{\sigma} (\vek) \notag\\
		&+ \sum_{\vek,\alpha,\beta}\psi_{\alpha}^{\dagger}(\vek)[A(k_x\sigma_x +k_y\sigma_y)]_{\alpha,\beta}\psi_{\beta}(\vek),\label{eq:H_TI}\\
		H_{\text{FI}} = & \sum_{\vek,\alpha,\beta}\psi_{\alpha}^{\dagger}(\vek)[m_z \sigma_z]_{\alpha,\beta}\psi_{\beta}(\vek),\label{eq:H_FI}\\
		H_{\text{SC}} = & \frac{1}{2}\sum_{\vek,\alpha,\beta}\{\psi_{\alpha}^{\dagger}(\vek)[\Delta(\theta)i\sigma_y ]_{\alpha,\beta}\psi_{\beta}^{\dagger}(-\vek) + \text{h.c.}\}\label{eq:H_SC}
	\end{align}
	\lina{as long as we consider positions far away from the interfaces at $x=0$ and $x=-d$ where $k_x$ is still a good quantum number.}
	Above, $\psi_{\sigma}^{(\dagger)}(\ve{k})$ annihilates (creates) an electron of momentum $\ve{k}$ and spin $\sigma$, and $(\sigma_x,\sigma_y,\sigma_z)$ are the Pauli matrices. To study the edge state at the QAHI/TSC interface, we construct the wave function in all three regions by taking into account all possible scattering processes of electron-like and hole-like particles from the superconducting region ($x>0$) to the non-superconducting non-magnetic region ($x<-d$), and vice versa (see \hyperlink{sec:MaterialsAndMethods}{Materials and Methods}) \cite{Lu_PTA_2018,Kashiwaya_RepProgPhys_2000}. In the limit $d\to\infty$, the chiral Majorana state at the QAHI/TSC interface ($x=0$) follows 
	the bound state energy dispersion \cite{Tanaka_PRL_2009}
	\begin{align}
		E_{\text{b}}(\theta) =
		\begin{cases}
			-\frac{\text{sgn}(m_z)|\Delta(\theta)| \mu \sin(\theta)}{\sqrt{m_z^2\cos^2(\theta)+\mu^2\sin^2(\theta)}} \text{ for } s\text{- and }d_{x^2-y^2}\text{-wave,}\\
			\phantom{+}\frac{\text{sgn}(\theta)|\Delta(\theta)| m_z \cos(\theta)}{\sqrt{m_z^2\cos^2(\theta)+\mu^2\sin^2(\theta)}} \text{ for }d_{xy}\text{-wave.}
		\end{cases}\label{eq:bound_state_energy}
	\end{align}
	\lina{The above dispersion describes the topologically non-trivial Majorana edge state. While the interface between a ferromagnet and a $d_{xy}$-wave superconductor holds two spin-degenerate zero energy modes that form Andreev bound states, the TI surface lifts the spin degeneracy leaving a single topologically non-trivial Majorana mode \cite{Linder_PRL_2010}. The dispersive nature of the Majorana mode was recently shown to have important implications for the optical conductivity: In the $s$-wave case, where the bound state dispersion is linear in the momentum $k_y$ at small incidence angles $\theta$, it was shown that a Cooper pair can absorb a photon and break up into two Majorana fermions which momentum $k_y$ along the edge both have the same sign \cite{He_PRL_2021}. Since a finite $k_y=k_{\text{F}}\sin(\theta)$ corresponds to a finite energy $E_{\text{b}}(\theta)$, the zero-temperature optical conductivity peaks at a finite energy, while it is zero at zero energy. \linak{Furthermore, it was recently predicted that the chirality of the Majorna edge can be directly probed via circularly polarized light \cite{Lu_PRB_2022}. The chirality and dispersive nature makes the Majorana edge mode} qualitatively distinct from the non-dispersive midgap Andreev bound states at the interface between a $d_{xy}$-wave superconductor and a ferromagnetic insulator \cite{Hu_PRL_1994,Sato_PRB_2011}. We also note that $E_{\text{b}}(\theta)$ is independent of the spin-orbit parameter $A$, which is true for all of the results presented in this manuscript.}
	
	From the wave function of the TSC, we construct the McMillan Green's function \cite{McMillan_PRL_1968,Kashiwaya_RepProgPhys_2000,Lu_PTA_2018}. From its retarded part, we numerically calculate the LDOS at the QAHI/TSC interface ($x=0$). As shown in Fig.~\ref{fig:01}(\textbf{B})-(\textbf{J}), the angle resolved LDOS has a single edge state that perfectly fits the bound state energy dispersion in \eqref{eq:bound_state_energy}. Since the LDOS is evaluated by performing the integral $\rho(x,\epsilon)=\int_{-\pi/2}^{\pi/2}d\theta\:\cos(\theta)\rho(x,\theta,\epsilon)$ over the angle resolved LDOS, small incidence angles $\theta$ give the dominant contributions to the LDOS. In the $s$-wave [Fig.~\ref{fig:01}(\textbf{B})-(\textbf{D})] and $d_{x^2-y^2}$-wave [Fig.~\ref{fig:01}(\textbf{E})-(\textbf{G})] TSC, the dispersion of the Majorana edge channel at small incidence angles $\theta$ flattens when increasing the magnetization in the QAHI \cite{Tanaka_PRL_2009,Linder_PRL_2010}. The nodes in the $d$-wave gap enhances the flatness of the \lina{bound state dispersion} of the $d_{x^2-y^2}$-wave TSC compared to the $s$-wave case. Compared to these two cases, the $d_{xy}$-wave TSC [Fig.~\ref{fig:01}(\textbf{H})-(\textbf{J})] is distinct in two ways: First, the steepness of the \lina{bound state dispersion} at small $\theta$ instead increases with increasing magnetization \cite{Linder_PRL_2010}. Secondly, the $d$-wave node is located at $\theta=0$ and thus gives a large contribution to the LDOS \cite{Hu_PRL_1994,Tanaka_PRL_1995}. We will show that the enhanced flatness of the \lina{bound state dispersion} of the $d$-wave TSC, together with contributions from the \lina{nodes in the superconducting gap}, give rise to distinct qualitative signatures in the optical conductivity through an enhanced signal below the 
	optical gap.
	
	\begin{figure*}[t!]
		\centering
		\includegraphics[width=\textwidth]{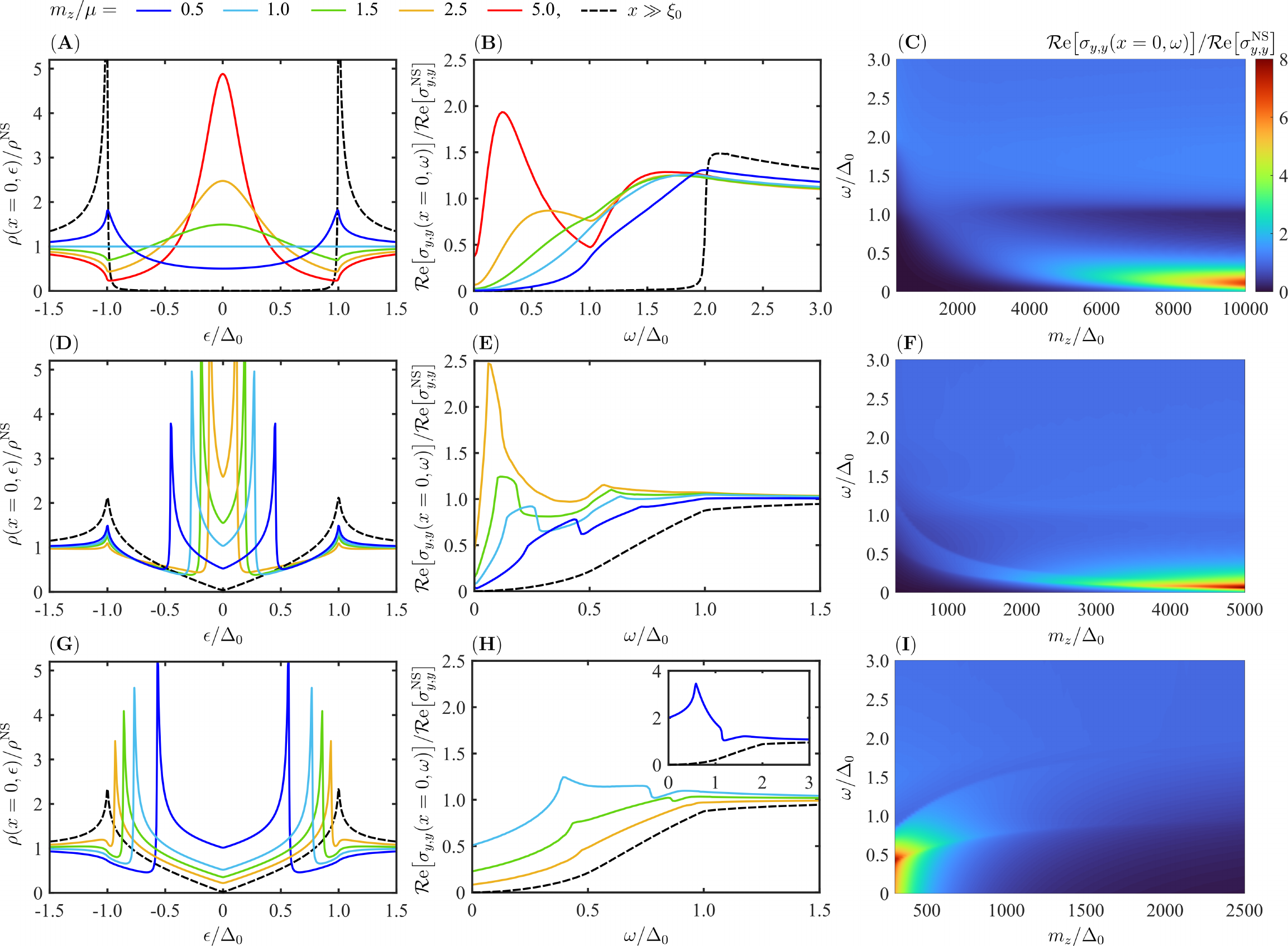}
		\caption{\textbf{Magnetization dependence of the local density of states and optical conductivity.}  For $s$-wave [panel (\textbf{A})-(\textbf{C})], $d_{x^2-y^2}$-wave [panel (\textbf{D})-(\textbf{F})], and $d_{xy}$-wave [panel (\textbf{G)}-(\textbf{I})] pairing, we consider the local density of states (LDOS) $\rho(x=0,\epsilon)$ [panel (\textbf{A}), (\textbf{D}), and (\textbf{G})] and the real part of the local optical conductivity $\mathcal{R}\text{e}[\sigma_{y,y}(x=0,\omega)]$ [panel (\textbf{B})-(\textbf{C}), (\textbf{E)}-(\textbf{F}), and (\textbf{H})-(\textbf{I})] normalized by their normal-state values for $\Delta_0 = 0$ at the QAHI/TSC interface ($x=0$). For comparison, the black dashed curves represent the LDOS and local optical conductivity far inside the TSC ($x\gg\xi_0$), where $\xi_0$ is the superconducting coherence length \lina{given by $k_{\text{F}}\xi_0=\mu/\Delta_0$}. For the $s$-wave TSC, the zero bias peak in the LDOS increases and the peak in the local optical conductivity increases and shifts towards zero energy  when increasing the magnetization $m_z$. For the $d_{x^2-y^2}$-wave ($d_{xy}$-wave) TSC, the peaks in the LDOS and optical conductivity increases and shifts towards zero energy with increasing (decreasing) magnetization. \lina{We consider $\Delta_0/\mu=10^{-3}\ll1$, and energies $\epsilon+i\delta$, where $\delta/\Delta_0=5\cdot10^{-3}$.}}
		\label{fig:02}
	\end{figure*}
	
	\begin{figure*}[t!]
		\centering
		\includegraphics[width=\textwidth]{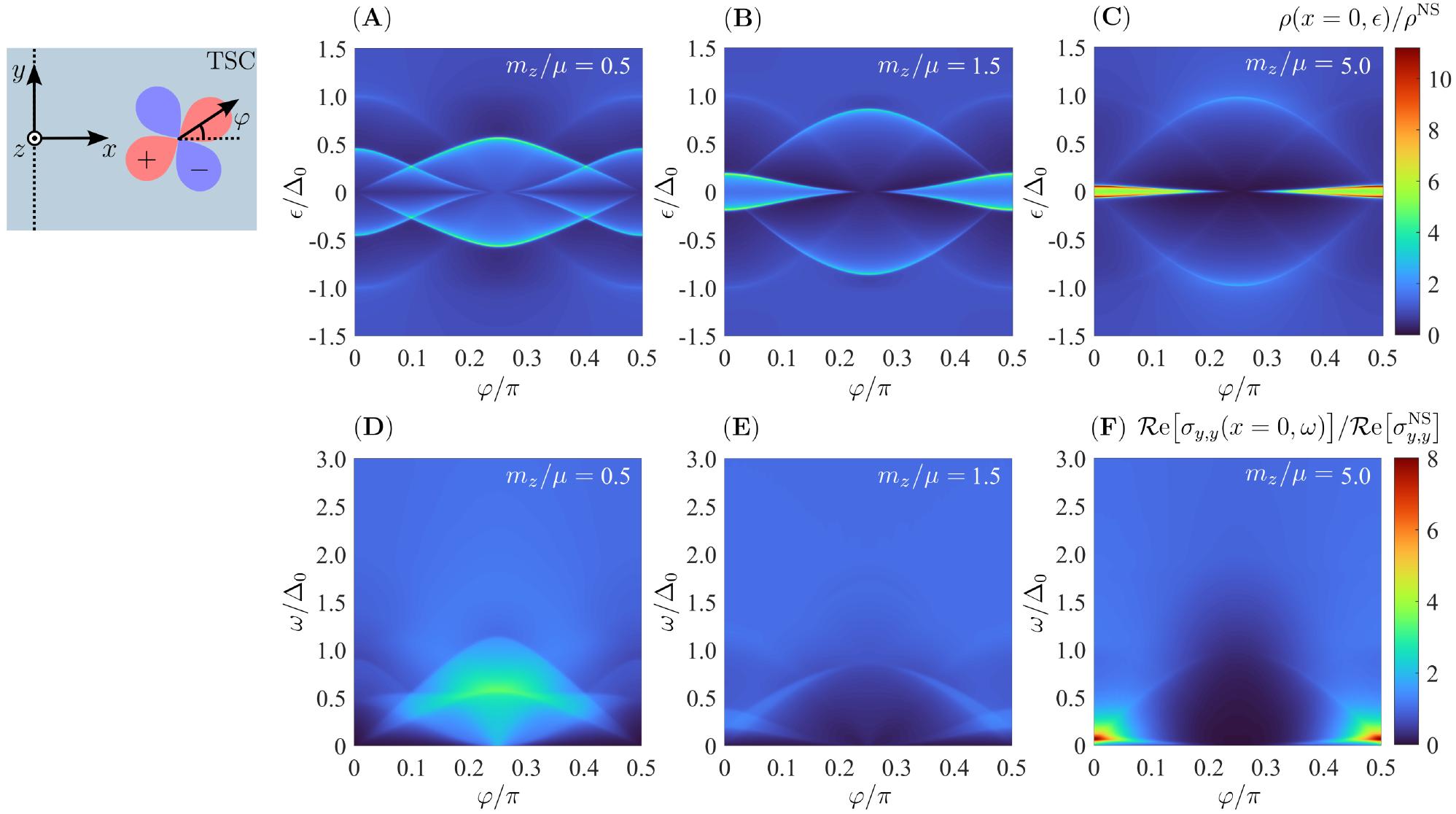}
		\caption{\textbf{Dependence of the local density of states and optical conductivity on the orientation of the $d$-wave nodes.} We consider the local density of states (LDOS) $\rho(x=0,\epsilon)$ [panel (\textbf{A})-(\textbf{C)}] and the real part of the local optical conductivity $\mathcal{R}\text{e}[\sigma_{y,y}(x=0,\omega)]$ [panel (\textbf{D})-(\textbf{F})] normalized by their normal-state values when $\Delta_0 = 0$ at the QAHI/TSC interface ($x=0$) under rotation of the $d$-wave lobe angle $\varphi$ with respect to the interface normal $\hat{x}$ (see schematic). We consider three values of the magnetization $m_z$ corresponding to Fig.~\ref{fig:01}(\textbf{E})-(\textbf{G}) ($\varphi = 0$) and Fig.~\ref{fig:01}(\textbf{H})-(\textbf{J}) ($\varphi=\pi/4$\lina{, $\Delta_0/\mu=10^{-3}\ll1$, and energies $\epsilon+i\delta$, where $\delta/\Delta_0=5\cdot10^{-3}$.} The LDOS and optical conductivity are either maximal at $\varphi=(0,\pi/2)$ ($d_{x^2-y^2}$-wave pairing) or at $\varphi=\pi/4$ ($d_{xy}$-wave pairing).}
		\label{fig:03}
	\end{figure*}
	
	\subsection*{Magnetization dependence of the optical conductivity}
	
	We use the Kubo formula \cite{Mahan_Book_2000} to numerically evaluate the local optical conductivity at the QAHI/TSC interface from the retarded and advanced McMillan Green's function \cite{McMillan_PRL_1968,Kashiwaya_RepProgPhys_2000,Lu_PTA_2018} of the TSC (see \hyperlink{sec:MaterialsAndMethods}{Materials and Methods} and Supplemental Information). We take into account contributions from both the Majorana edge state and energy states above the superconducting gap. This is of particular importance in the $d$-wave case, where the nodes in the superconducting gap [see Fig.~\ref{fig:01}(\textbf{E})-(\textbf{J})] contribute to the optical conductivity at energies well below the optical gap $\omega=2\Delta_0$. 
	\linajk{The optical signal can be measured from above via MIM. During the measurement, the MIM tip  is placed above the sample and moved along the $x$ axis across the QAHI/TSC interface as indicated by the white dotted line in Fig.~\ref{fig:01}(\textbf{A}) \cite{Wang_PRB_2023}. Due to its energy gap, the ferromagnetic insulator is transparent at the energy scale relevant for observing the response of the Majorana edge states. The superconductor can contribute with an additional signals due to its finite skin depth, and due to the nodes and sensitivity to impurities in the $d$-wave case. It is therefore preferable to consider a thin and clean superconducting film. Further details regarding the experimental realization are discussed in the Supplemental Information.} 
	We first consider the local optical conductivity exactly at the QAHI/TSC interface ($x=0$) at zero temperature, and address the decay of the edge state 
	inside the TSC, as well as finite temperatures later on.
	
	To understand the results for the optical conductivity, we first consider how the LDOS depends on the magnetization in the QAHI. As already hinted by the angle resolved LDOS in Fig.~\ref{fig:01}, the edge state gives rise to peaks in the LDOS for energies $|\epsilon|<\Delta_0$. In the $s$-wave case [Fig.~\ref{fig:02}(\textbf{A})], a single peak with maxima at zero energy develops with increasing magnetization. \lina{While in the $s$-wave case [Fig.~\ref{fig:01}(\textbf{B})-(\textbf{D})], the bound state dispersion only crosses zero energy once, the nodes in the $d$-wave case forces the bound state dispersion to cross zero energy three times [Fig.~\ref{fig:01}(\textbf{E})-(\textbf{J})]. The increased curvature of the bound state dispersion caused by the $d$-wave nodes results in two LDOS peaks with maxima at finite energy in the $d_{x^2-y^2}$-wave [Fig.~\ref{fig:02}(\textbf{D})] and $d_{xy}$-wave [Fig.~\ref{fig:02}(\textbf{G})] case}. These shift towards lower energies and increase in height with increasing (decreasing) magnetization in the $d_{x^2-y^2}$-wave ($d_{xy}$-wave) case\lina{, thus reflecting the behavior of the bound state dispersion}. As the bound state dispersion becomes flatter, the contribution from energies $|\epsilon|>\Delta(\theta)$ is suppressed [see Fig.~\ref{fig:01}(\textbf{D}), (\textbf{G}), and (\textbf{H}) compared to Fig.~\ref{fig:01}(\textbf{B}), (\textbf{E}), and (\textbf{J}), respectively]. Thus, as the peaks resulting from the edge state increase in height, the \lina{states above the superconducting gap} contribute less to the LDOS.
	
	The edge state similarly results in a peak in the local optical conductivity at a finite energy below the optical gap $0<\omega<2\Delta_0$. \lina{This was previously shown to be a distinct qualitative feature resulting from the dispersion of the chiral Majorana mode \cite{He_PRL_2021}.} In the $s$-wave [Fig.~\ref{fig:02}(\textbf{B})-(\textbf{C})] and $d_{x^2-y^2}$-wave [Fig.~\ref{fig:02}(\textbf{E})-(\textbf{F})] case, this peak increases in height and shifts towards lower energies as the magnetization increases. The peak height increases more rapidly with increasing magnetization in the $d_{x^2-y^2}$-wave case\lina{, as explained by the flatness of the dispersion of the edge state}.  
	At $\omega=0$, the small but finite value of the optical conductivity results from having a finite imaginary part of the energy $\epsilon+i\delta$, but vanishes in the limit $\delta\to0$ (see Supplemental Information) since the Majorana mode can  contribute to the optical conductivity only at finite energies \cite{He_PRL_2021}.
	\lina{Our results for the $s$-wave case are thus fully consistent with those presented in Ref.~\cite{He_PRL_2021}.}
	In the case of $d_{xy}$-wave pairing [Fig.~\ref{fig:02}~(\textbf{H})-(\textbf{I})], the optical conductivity is finite at zero energy even in the limit $\delta\to0$. The finite value in the $d_{xy}$-wave case arises from contributions from the node at $\theta=0$. Contrary to the $s$-wave and $d_{x^2-y^2}$-wave cases, the peak value in the $d_{xy}$-wave case decreases and shifts towards higher energies as the magnetization increases \lina{due to the opposite behavior of the bound state dispersion in response to an increase in the magnetization $m_z$ in the $d_{xy}$-wave case.} Thus, large peaks in the local optical conductivity can be achieved at weaker magnetization.
	
	\subsection*{Optical conductivity for general orientations of the $d$-wave nodes}
	
	We next consider a general orientation of the $d$-wave nodes by studying the dependence of the LDOS and local optical conductivity on the angle $\varphi$ of the positive $d$-wave lobe with respect to the interface normal $\ve{x}$. The LDOS contains two pairs of peaks with maximum value at $\varphi=(0,\pi/2)$ and $\varphi=\pi/4$, respectively [Fig.~\ref{fig:03}(\textbf{A})-(\textbf{C})]. At low magnetization, the peaks at $\varphi=(0,\pi/2)$ are well separated and the peaks at $\varphi=\pi/4$ dominate [panel~(\textbf{A})]. As we increase the magnetization, the peaks at $\varphi=\pi/4$ decrease and shifts towards higher energies, while the peaks at $\varphi=(0,\pi/2)$ shifts towards zero energy and increase in height [panel~(\textbf{C})]. As a result, the optical conductivity [Fig.~\ref{fig:03}(\textbf{D})-(\textbf{F})] has a peak at $\varphi=\pi/4$ ($d_{xy}$-wave pairing) at low magnetization [panel (\textbf{D})] and at $\varphi=(0,\pi/2)$ ($d_{x^2-y^2}$-wave pairing) at higher magnetization [panel (\textbf{F})], and never for intermediate angles.
	
	\subsection*{Position dependence of the optical conductivity}
	
\begin{figure}[htb]
		\centering
		\includegraphics[width=\columnwidth]{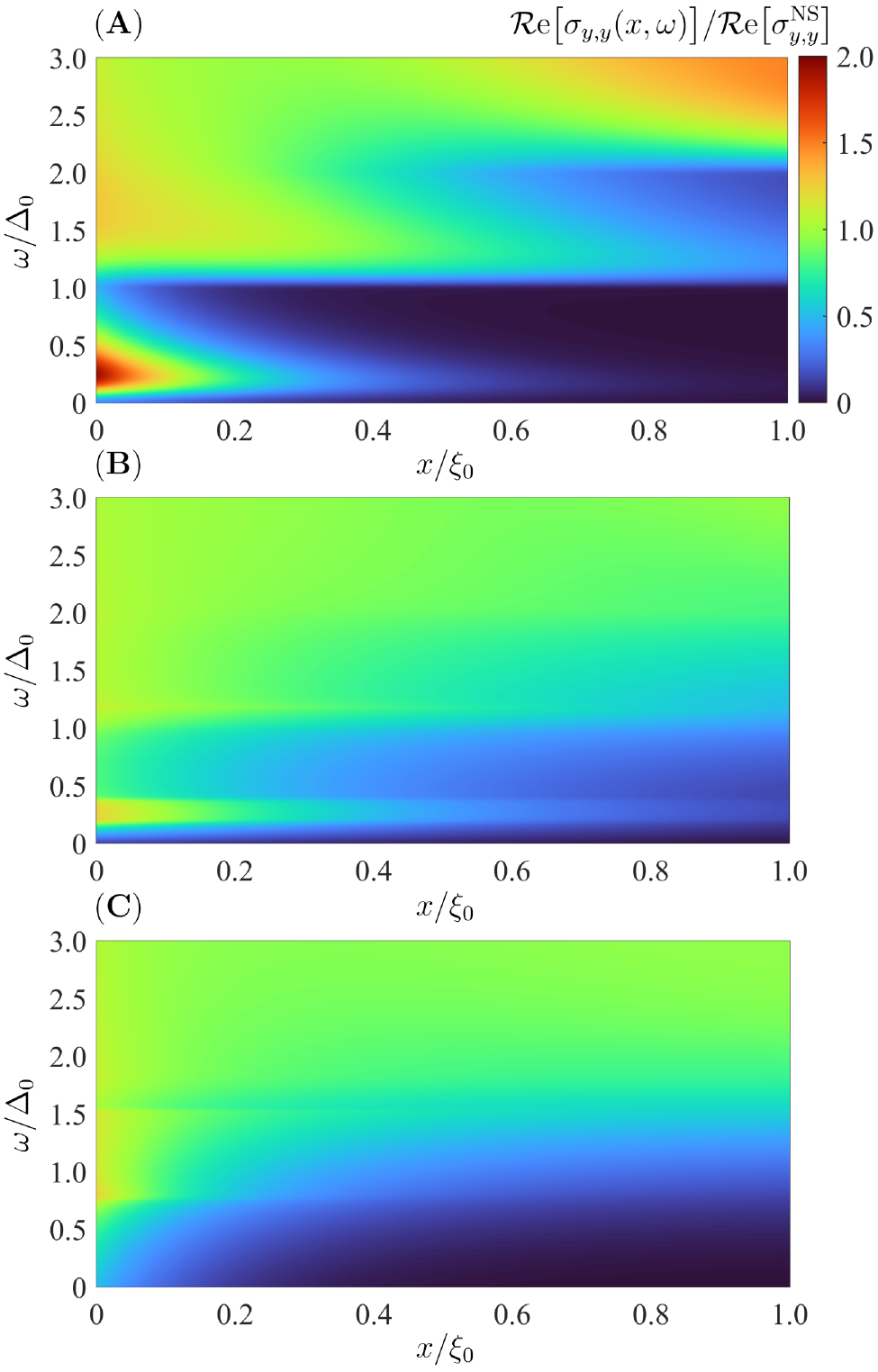}
		\caption{\textbf{The optical conductivity as a function of the distance from the interface.} We consider the real part of the optical conductivity $\mathcal{R}\text{e}[\sigma_{y,y}(x,\omega)]$ normalized by its normal-state value when $\Delta_0 = 0$ as a function of the distance~$x$ from the QAHI/TSC interface for $s$-wave pairing [panel (\textbf{A})], $d_{x^2-y^2}$-wave pairing [panel (\textbf{B})], and $d_{xy}$-wave pairing [panel (\textbf{C})]. The corresponding magnetizations are given by \lina{$m_z/\mu=5.0$, $1.5$ and $1.0$, respectively. We consider $\Delta_0/\mu=10^{-3}\ll1$, and energies $\epsilon+i\delta$, where $\delta/\Delta_0=5\cdot10^{-3}$.} The features below the optical gap $\omega<2\Delta_0$ decay monotonically over a length scale comparable to the superconducting coherence length $\xi_0$.}
		\label{fig:04}
	\end{figure}

	So far, we have considered the optical conductivity exactly at the QAHI/TSC interface ($x=0$). \lina{In experiments, the optical conductivity would however be measured by placing the MIM tip above the sample and moving it along the $x$ axis across the QAHI/TSC interface, as indicated in Fig.~\ref{fig:01}(\textbf{A}) \cite{Wang_PRB_2023}.} We now consider how the optical conductivity varies away from the interface at a fixed magnetization [see Fig.~\ref{fig:04}]. In both the $s$-wave and $d$-wave TSC, the features below the optical gap ($\omega<2\Delta_0$) decay over a length scale comparable to the superconducting coherence length $\xi_0$, \lina{here given by $k_{\text{F}}\xi_0 = \mu/\Delta_0$}. Above the optical gap ($\omega>2\Delta_0$), the \lina{behavior expected for $x\gg\xi_0$} is restored when moving away from the interface. In the $s$-wave case [Fig.~\ref{fig:04}(\textbf{A})], the shifting of the gap edge towards lower energies leads to a characteristic double peak feature at intermediate distances from the interface where the low energy peak is still present and the original optical gap is partly restored. 
	This feature is less prominent in the case of $d$-wave pairing [Fig.~\ref{fig:04}(\textbf{B})-(\textbf{C})] due to the absence of a hard gap. 
	\linajk{When considering a finite spot size for the MIM measurement, the local optical conductivity in Fig.~\ref{fig:04} is averaged over the spot size. The position-averaged optical conductivity is presented in the Supplemental Information.} \linat{We find that while such an averaging reduces the peak associated with the Majorana edge mode compared to its local value at the interface, the peak remains measurable for a realistic spot size.}
	\lina{While our model does not allow for determining the $z$ axis dependence of the optical conductivity, the results are not expected to be significantly altered if the Majorana edge mode is not perfectly confined at the TI surface as long as the optical conductivity is measured from above, effectively summing up contributions from different values of $z$.}
	
	\begin{figure*}[htb]
		\centering
		\includegraphics[width=\textwidth]{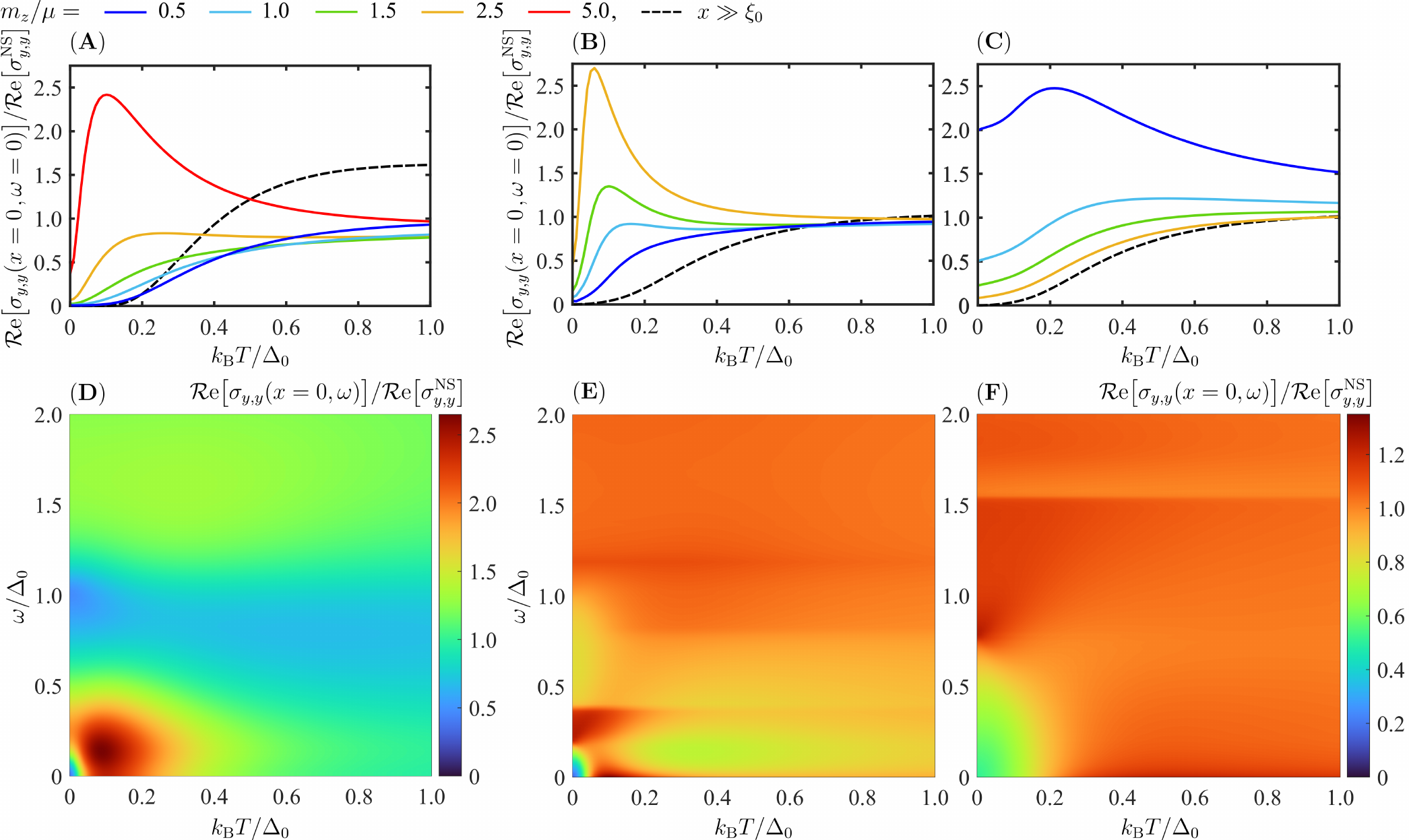}
		\caption{\textbf{Temperature dependence of the optical conductivity.} We consider how the real part of the local optical conductivity $\mathcal{R}\text{e}[\sigma_{y,y}(x=0,\omega)]$ normalized by its normal-state value when $\Delta_0 = 0$ at the QAHI/TSC interface ($x=0$) varies as we increase the temperature $T$ for zero energy ($\omega=0$) [panel (\textbf{A})-(\textbf{C})] and for energies below the optical gap ($\omega<2\Delta_0$) [panel (\textbf{D})-(\textbf{F})]. From left to right, we consider $s$-wave [panel~(\textbf{A}) and~(\textbf{D})], $d_{x^2-y^2}$-wave [panel (\textbf{B}) and (\textbf{E})], and $d_{xy}$-wave [panel (\textbf{C}) and (\textbf{F})] pairing. The results in panel~(\textbf{D})-(\textbf{F}) correspond to magnetizations given by \lina{$m_z/\mu=5.0$, $1.5$ and $1.0$, respectively. For comparison, the black dashed curves represent the local optical conductivity far inside the TSC ($x\gg\xi_0$). We consider $\Delta_0/\mu=10^{-3}\ll1$, and energies $\epsilon+i\delta$, where $\delta/\Delta_0=5\cdot10^{-3}$.} In the $s$-wave and $d_{x^2-y^2}$-wave ($d_{xy}$-wave) case, the temperature dependence becomes non-monotonic when the magnetization increases (decreases).}
		\label{fig:05}
	\end{figure*}
	
	\subsection*{Temperature dependence of the optical conductivity}
	
	We finally consider how the optical conductivity behaves as the temperature increases (see Fig.~\ref{fig:05}). Considering the limit $\omega\to0$ [Fig.~\ref{fig:05}(\textbf{A})-(\textbf{C})], we find that while the optical conductivity always increases for $k_{\text{B}}T\ll\Delta_0$, the behavior at higher temperatures can be non-monotonic. For the $s$-wave and $d_{x^2-y^2}$-wave TSC [panel~(\textbf{A}) and~(\textbf{B}), respectively], the non-monotonic temperature dependence appears when the magnetization increases. For the $d_{xy}$-wave TSC [panel~(\textbf{C})], it instead appears when the magnetization decreases.
	This can be understood by studying the expression for the real part of the local optical conductivity [\eqref{eq:oc} in Materials and Methods] in the limit $\omega\to0$ where it essentially consists of an energy integral over a temperature dependent factor 
	\begin{align}
		\lim_{\omega\to0}\frac{f_{\text{FD}}(\epsilon-\omega)-f_{\text{FD}}(\epsilon)}{\omega}=\frac{1}{k_{\text{B}}T}e^{\epsilon/k_{\text{B}}T}[f_{\text{FD}}(\epsilon)]^2,\label{eq:Tdep}
	\end{align}
	and a product of Green's functions. Above, $f_{\text{FD}}(\epsilon)$ is the Fermi-Dirac distribution and $k_{\text{B}}$ is the Boltzmann constant. The product of Green's functions can be assumed temperature independent when the maximum value of the superconducting gap $\Delta_0$ proximity-induced \lina{onto the TI surface} is much smaller than in the parent SC. In this case, the parent SC, and thus the proximity-induced $\Delta_0$, are nearly unaffected by the $k_{\text{B}}T/\Delta_0$ considered. 
	The Green's function products are symmetric in the energy $\epsilon$ and have coherence peaks at $\epsilon=\pm\Delta_0$. At the QAHI/TSC interface, they contain additional features for energies $0<\epsilon<\Delta_0$ similar to the peaks in the zero-temperature optical conductivity for $0<\omega<2\Delta_0$ resulting from the edge state [see Fig.~\ref{fig:02}(\textbf{B}), (\textbf{E}), and (\textbf{H})]. 
	The temperature dependent factor in \eqref{eq:Tdep} is peaked at zero energy with height $1/4k_{\text{B}}T$ and width $2\text{ln}(3+2\sqrt{2})k_{\text{B}}T$ at half of the peak height. It thus determines which features of the Green's function products are included through the broadening of the peak as $T$ increases. A non-monotonic temperature dependence is possible when the features in the Green's function products resulting from the edge state enters the peak width of the temperature dependent factor at small but finite temperatures. A finite magnetization additionally alters the height of the coherence peaks at the QAHI/TSC interface, so that the optical conductivity does not necessarily approach \lina{the result for $x\gg\xi_0$} at higher temperatures.
	
	In Fig.~\ref{fig:05}(\textbf{D})-(\textbf{F}), we consider the temperature dependence for all energies $\omega$ below the optical gap ($\omega<2\Delta_0$) at a given magnetization. When the behavior of the $s$-wave TSC [panel~(\textbf{D})] is non-monotonic in the limit $\omega\to0$, the optical conductivity peaks at finite energy and finite temperature. For the $d_{x^2-y^2}$-wave TSC [panel~(\textbf{E})], the behavior is even more non-trivial with a saddle point separating peaks at finite energy and finite temperature. While the optical conductivity of the $s$-wave and $d_{x^2-y^2}$-wave TSC approaches a small value (zero in the limit $\delta\to0$, see Supplemental Information) when $(k_{\text{B}}T,\omega)\to0$, the optical conductivity of the $d_{xy}$-wave TSC [panel~(\textbf{F})] approaches a finite value due to the nodal states. This causes the features of the $d_{xy}$-wave TSC to be smeared out.
	
	\section*{Discussion}
	
	The recent advances in microwave impedance microscopy of topological edge states \cite{Allen_PNAS_2019,Wang_PRB_2023} open a promising new avenue for probing the local optical conductivity of the Majorana edge state at the QAHI/TSC interface \cite{He_PRL_2021}. 
	We have presented a series of \lina{features, characteristic of the chiral Majorana mode due to its bound state dispersion,} accessible via this technique. The qualitative behavior of the local optical conductivity as a function of energy, magnetization and temperature depend on the symmetry of the superconducting order parameter through contributions from additional nodal states that enhance the flatness of the bound state dispersion. Qualitatively distinct behavior is expected when comparing the $s$-wave and $d_{x^2-y^2}$-wave cases to the $d_{xy}$-wave case. When increasing or decreasing the magnetization, respectively, signatures of the Majorana edge state appear in the form of a peak below the optical gap and a non-monotonic temperature dependence. Distinct and tunable signatures also appear in the LDOS accessible via scanning tunneling spectroscopy measurements 
	\cite{Menard_NatCommun_2017,Palacio-Morales_SciAdv_2019,Wang_Science_2020,Kezilebieke_Nature_2020}.
	
	\lina{While the Majorana edge mode is topologically protected from \linak{backward}-scattering due to} \linak{its chiral nature} \lina{\cite{Chung_PRB_2011}, our results are more relevant in experimentally achievable \cite{Zareapour_NatCommun_2012} clean systems since the $d$-wave pairing is not protected by Anderson's theorem \cite{Anderson_JPhysChemSol_1959} and thus not robust to disorder. The chiral Majorana edge mode considered here is dispersive and one-dimensional and runs along the entire length of the interface. \linak{It can therefore be distinguished from} trivial Andreev bound states \linak{close to impurities in the $d$-wave superconductor. These} may give rise to non-dispersive zero-energy states that would show up as a peak in STM measurements near the localized impurity \cite{Jack_NatRevPhys_2021}, \lina{but} cannot provide the conductive channel along the interface that gives rise to the finite optical conductivity. The optical conductivity signal can moreover be distinguised from that of flat Andreev bound states at the interface between a $d_{xy}$-wave SC and a FI, since these are non-dispersive zero-energy states \cite{Hu_PRL_1994,Linder_PRL_2010}. It is the Majorana edge mode's dispersive nature that that gives rise to the characteristic optical conductivity peak at finite energy \cite{He_PRL_2021}.}
	
	Note that we assume the limit of small tip in LDOS and small spot of the optical excitation in $\sigma_{y,y}(\omega)$. The former is usually applicable for the scanning tunnelling spectroscopy, while the latter is not trivial. Recent advances of the \lina{MIM} experiment achieved the ultrahigh spatial resolution of 5 nm \cite{doi:10.1126/sciadv.abd1919}, but this is still nearly 10 times larger than the lattice constant. Therefore, the momentum transfer associated with the optical transition is of the order of the inverse of 5 nm. However, we are interested in the low energy region below the superconducting gap, and the relevant length scale is the coherence length of the superconductivity, which is larger than 5 nm.
	
	\linajk{In addition to the position averaging of the optical signal over the spot size, contributions to the optical signal can be picked up from the superconductor deposited on top of the TI surface due to its finite skin depth, and in the $d$-wave case, due to nodal states and disorder. However, $d$-wave superconductivity in Bi$_2$Sr$_2$CaCu$_2$O$_{8-\delta}$ has been measured down to the monolayer limit with an approximate thickness of 2~nm \cite{Yu_Nature_2019}, well below the zero-frequency zero-temperature skin depth which is of the order of 100~nm. Moreover, taking into account contributions from a disordered $d$-wave superconductor on top of the TI, the conductance peak resulting from the edge states give the dominating contribution to the total signal for a superconducting film up to the order of hundred atomic layers, and a measurable signal up to the order of $10^4$ atomic layers relevant to, \textit{e.g.}, the experiments in Ref.~\cite{Zareapour_NatCommun_2012}. Further details on how our calculation is relevant to experiments is discussed in the Supplemental Information}. 
	
	The advantage associated with the qualitative nature of these results lies in their tunability and robustness with respect to measurement noise. 
	Previous measurements using scanning tunneling microscopy relied on quantitative measures through, \textit{e.g.}, a quantized conductance \cite{Chung_PRB_2011,Wang_PRB_2015,Lian_PRB_2016}. Since an exact quantization can only be achieved in theory, it is challenging to distinguish whether an apparent quantization measured in the lab is of trivial or topological origin \cite{Ji_PRL_2018,Huang_PRB_2018,Kayyalha_Science_2020}. 
	The rich behavior of the local optical conductivity and density of states presented here -- distinct between the different symmetries of the superconducting order parameter -- lays a broader foundation to account for non-trivial behavior through magnetization and temperature 
	dependencies specific to the Majorana edge mode.
	
	\matmethods{
		\subsection*{Wave functions}
		\hypertarget{sec:MaterialsAndMethods}{To} construct the McMillan Green's function \cite{McMillan_PRL_1968,Kashiwaya_RepProgPhys_2000,Lu_PTA_2018,Tanaka_PRL_2009,Linder_PRL_2010}, we first construct the ordinary wave functions $\Psi_j(x,y)$ of the four possible scattering processes ($j=1,2,3,4$) of an electron-like and a hole-like particle scattering from the non-superconducting non-magnetic region ($x<-d$) to the superconducting region ($x>0$), and vice versa. We also construct the conjugated wave functions $\widetilde{\Psi}_j(x,y)$ of the reverse scattering processes. The wave functions are constructed from the eigenvectors obtained by diagonalizing the Hamiltonian in \eqref{eq:H_TI} - \eqref{eq:H_SC}. Their momenta along $\hat{x}$ are derived from the corresponding eigenenergies. We assume the momentum $k_y=(\mu/A)\sin(\theta)$ along the interface to be conserved during the scattering process so that 
		\begin{align}
			\parwidetilde{\Psi}(x,y)=\parwidetilde{\Psi}(x)e^{\horpm ik_y y}    
		\end{align}
		In the superconducting region ($x>0$), the wave functions are given by
		\begin{align}
			\parwidetilde{\Psi}_j (x)=&\parwidetilde{\Psi}_j^{\text{SC,in}}(x)
			+\parwidetilde{a}_j\parwidetilde{\Psi}^{\text{SC}}_{\text{eR}}e^{ik_x^{\text{SCe}\horpm}x}+\parwidetilde{b}_j\parwidetilde{\Psi}_{\text{hR}}^{\text{SC}}e^{-ik_x^{\text{SCh}\hormp}x},
		\end{align}
		with incoming wave functions
		\begin{align}
			\parwidetilde{\Psi}_1^{\text{SC,in}}(x)=\parwidetilde{\Psi}_2^{\text{SC,in}}(x)=0,\:\:
			&\parwidetilde{\Psi}_3^{\text{SC,in}}(x)=\parwidetilde{\Psi}_{\text{eL}}^{\text{SC}}e^{-ik_x^{\text{SCe}\hormp}x},\notag\\
			&\parwidetilde{\Psi}_4^{\text{SC,in}}(x)=\parwidetilde{\Psi}_{\text{hL}}^{\text{SC}}e^{ik_x^{\text{SCh}\horpm}x},
		\end{align}
		where $a_j$ and $b_j$ are coefficients, the wave vectors are given by
		\begin{align}
			\Psi_{\text{eR(L)}}^{\text{SC}} &= [1\:\:\:\horpm e^{\horpm i\theta}\:\:\:\hormp\Gamma_{\horpm}e^{\horpm i\theta}\:\:\:\Gamma_{\horpm}]^T,\label{eq:vecSC1}\\
			\Psi_{\text{hL(R)}}^{\text{SC}} &= [\Gamma_{\horpm}\:\:\:\horpm\Gamma_{\horpm}e^{\horpm i\theta}\:\:\:\hormp e^{\horpm i\theta}\:\:\:1]^T,\\
			\widetilde{\Psi}_{\text{eR(L)}}^{\text{SC}} &= [1\:\:\:\hormp e^{\horpm i\theta}\:\:\:\horpm \Gamma_{\hormp}e^{\horpm i\theta}\:\:\:\Gamma_{\hormp}]^T,\\
			\widetilde{\Psi}_{\text{hL(R)}}^{\text{SC}} &= [\Gamma_{\hormp}\:\:\:\hormp\Gamma_{\hormp}e^{\horpm i\theta}\:\:\:\horpm e^{\horpm i\theta}\:\:\:1]^T,\label{eq:vecSC4}
		\end{align}
		and
		\begin{align}
			\Gamma_{\horpm} &=\frac{\Delta_0\cos\{2[\theta\hormp\varphi]\}}{\epsilon+\sqrt{\epsilon^2-|\Delta_0\cos\{2[\theta\hormp\varphi]\}|^2}}.
		\end{align}
		The momenta $k_x^{\text{SCe(h)}\pm}$ of electron-like (hole-like) particles are given by
		\begin{align}
			A k_x^{\text{SCe(h)}\pm} &=\sqrt{\left\{\mu\horpm\sqrt{\epsilon^2-\left|\Delta_0\cos\left[2\left(\theta\mp \varphi\right)\right]\right|^2}\right\}^2-A^2k_y^2}.
		\end{align}
		In this region, we have assumed $0<(\epsilon,\Delta_0)\ll\mu$.
		In the ferromagnetic region ($-d<x<0$), the wave functions are given by 
		\begin{align}
			\parwidetilde{\Psi}_{j}^{\text{FI}}=&\left(\parwidetilde{c}_j \parwidetilde{\Psi}_{\text{eL}}^{\text{FI}}
			+\parwidetilde{d}_j \parwidetilde{\Psi}_{\text{hR}}^{\text{FI}}\right)e^{\kappa_x^{\text{FI}}x}\notag\\
			+&\left(\parwidetilde{e}_j \Psi_{\text{eR}}^{\text{FI}}
			+\parwidetilde{f}_j \parwidetilde{\Psi}_{\text{hL}}^{\text{FI}}\right)e^{-\kappa_x^{\text{FI}}x}
			,
		\end{align}
		where $c_j$, $d_j$, $e_j$ and $f_j$ are coefficients, the wave vectors are given by
		\begin{align}
			\Psi_{\text{eR(L)}}^{\text{FI}}&=[\horpm i\gamma^{\horpm1}\:\:\:\:1\:\:\:\:0\:\:\:\:0]^T,\\
			\Psi_{\text{hL(R)}}^{\text{FI}}&=[0\:\:\:\:0\:\:\hormp i\gamma^{\hormp1}\:\:\:\:1]^T,\\
			\widetilde{\Psi}_{\text{eR(L)}}^{\text{FI}}&=[\hormp i\gamma^{\horpm1}\:\:\:\:1\:\:\:\:0\:\:\:\:0]^T,\\
			\widetilde{\Psi}_{\text{hL(R)}}^{\text{FI}}&=[0\:\:\:\:0\:\:\horpm i\gamma^{\hormp1}\:\:\:\:1]^T,
		\end{align}
		with $\gamma = -A(\kappa_x^{\text{FI}}-k_y)/m_z$, and the momentum is given by $A\kappa_x^{\text{FI}}=\sqrt{m_z^2+(Ak_y)^2}$.
		In this region, we have assumed $\mu=0$ and $0<\epsilon\ll|m_z|$.
		In the non-superconducting non-magnetic region ($x<-d$), the wave functions are given by 
		\begin{align}
			\parwidetilde{\Psi}_j(x)=\parwidetilde{\Psi}_j^{\text{N,in}}(x)+\parwidetilde{g}_j\parwidetilde{\Psi}_{\text{eL}}^{\text{N}}e^{-ik_x^{\text{N}}x}+\parwidetilde{h}_j\parwidetilde{\Psi}_{\text{hL}}^{\text{N}}e^{ik_x^{\text{N}}x},
		\end{align}
		with incoming wave functions
		\begin{align}
			&\parwidetilde{\Psi}_1^{\text{N,in}}(x)=\parwidetilde{\Psi}_{\text{eR}}^{\text{N}}e^{ik_x^{\text{N}}x},\notag\\
			&\:\parwidetilde{\Psi}_2^{\text{N,in}}(x)=\parwidetilde{\Psi}_{\text{hR}}^{\text{N}}e^{-ik_x^{\text{N}}x},
			\:\:\parwidetilde{\Psi}_3^{\text{N,in}}(x)=\parwidetilde{\Psi}_4^{\text{N,in}}=0,
		\end{align}
		where $g_j$ and $h_j$ are coefficients, and the wave vectors are obtained by setting $\Gamma_{\horpm}=0$ in \eqref{eq:vecSC1} - \eqref{eq:vecSC4}.
		The momentum is given by $Ak_x^{\text{N}}=\sqrt{\mu^2-A^2k_y^2}$.
		The coefficients are evaluated by imposing continuity of the ordinary and conjugated wave function at $x=-d$ and $x=0$. 
		
		\subsection*{The McMillan Green's function}
		
		From the ordinary and conjugated wave functions, we can construct the retarded McMillan Green's function \cite{McMillan_PRL_1968,Kashiwaya_RepProgPhys_2000,Lu_PTA_2018}
		\begin{align}
			G^{\text{R}} (x,x') =
			\begin{cases}
				&\phantom{+}\alpha_1 \Psi_1(x)\widetilde{\Psi}_3^T(x')
				+\alpha_2 \Psi_1(x)\widetilde{\Psi}_4^T(x')\\
				&+\alpha_3 \Psi_2(x)\widetilde{\Psi}_3^T(x')
				+\alpha_4 \Psi_2(x)\widetilde{\Psi}_4^T(x'),\:x>x',\\
				&\phantom{+}\beta_1 \Psi_3(x)\widetilde{\Psi}_1^T(x')
				+\beta_2 \Psi_4(x)\widetilde{\Psi}_1^T(x')\\
				&+\beta_3 \Psi_3(x)\widetilde{\Psi}_2^T(x')
				+\beta_4 \Psi_4(x)\widetilde{\Psi}_2^T(x'),\:x<x'.
			\end{cases}
		\end{align}
		The coefficients $\alpha_j$ and $\beta_j$ are obtained from the boundary condition
		\begin{align}
			\lim_{\delta\to0^+}G^{\text{R}}(x'+\delta,x')-G^{\text{R}}(x'-\delta,x') = -i\tau_0\sigma_x/A,
		\end{align}
		where $\tau_0$ is the unit matrix in Nambu space.
		The resulting retarded Green's function is a $4\times4$ matrix 
		\begin{align}
			&G^{\text{R}}(x,x',\theta,\epsilon+i\delta)=
			\begin{pmatrix}
				g^{\text{R}}(x,x',\theta,\epsilon+i\delta) & f^{\text{R}}(x,x',\theta,\epsilon+i\delta) \\
				\underline{f^{\text{R}}}(x,x',\theta,\epsilon+i\delta) & \underline{g^{\text{R}}}(x,x',\theta,\epsilon+i\delta)
			\end{pmatrix}
		\end{align}
		written in terms of $2\times2$ ordinary and anomalous retarded Green's functions (see Supplemental Information for analytic expressions). The incidence angle $\theta$ and energy $\epsilon+i\delta$, where $\delta>0$ is a small parameter, was left out until now for simplicity of notation. We neglect terms that oscillates over a length scale much smaller than the superconducting coherence length.
		Since we assumed $\epsilon>0$ when constructing the wave functions, negative energies are accessed from the lower elements via the relation
		\begin{align}
			g^{\text{R}}(x,x',\theta,-\epsilon+i\delta)&=-[\underline{g^{\text{R}}}(x,x',-\theta,\epsilon+i\delta)]^*,\\
			f^{\text{R}}(x,x',\theta,-\epsilon+i\delta)&=-[\underline{f^{\text{R}}}(x,x',-\theta,\epsilon+i\delta)]^*.
		\end{align}
		To find the advanced Green's function, we similarly evaluate
		\begin{align}
			g^{\text{A}}_{\alpha,\beta}(x,x',\theta,\epsilon-i\delta)& = \left[g^{\text{R}}_{\beta,\alpha}(x',x,\theta,\epsilon+i\delta)\right]^*,\\
			f^{\text{A}}_{\alpha,\beta}(x,x',\theta,\epsilon-i\delta) &= -f^{\text{R}}_{\beta,\alpha}(x',x,-\theta,-\epsilon+i\delta).
		\end{align}

		\subsection*{The local density of states and optical conductivity}
		
		The LDOS is given by 
		\begin{align}
			\rho(x,\epsilon)=-\frac{1}{\pi}\int_{-\pi/2}^{\pi/2}d\theta\:\cos(\theta)\Imag\{\text{Tr}[g^{\text{R}}(x,x,\theta,\epsilon)]\}.
		\end{align}
		In the normal-state ($\Delta_0 = 0$), $\rho(x,\epsilon)=1/A$ is constant.
		The real part of the optical conductivity (see Supplemental Information for its derivation) is given by
		\begin{align}
			\Real\bigg[&\sigma_{i,j}(x\neq x',\omega+i\delta)\bigg]=\notag\\
			\Real\bigg[&\frac{e^2\mu A}{8\omega}\sum_{\alpha,\beta}\sum_{\alpha',\beta'}\sigma_{\alpha,\beta}^i\sigma_{\alpha',\beta'}^j
			\frac{1}{\pi}\int_{-\infty}^{\infty}d\epsilon\:[f_{\text{FD}}(\epsilon)-f_{\text{FD}}(\epsilon-\omega)]\notag\\
			&\frac{1}{\pi}\int_{-\pi/2}^{\pi/2}d\theta\:\cos(\theta)\frac{1}{\pi}\int_{-\pi/2}^{\pi/2}d\theta'\:\cos(\theta')\notag\\
			\big\{\big[&g^{\text{R}}_{\beta,\alpha'}(x,x',\theta,\epsilon)-g^{\text{A}}_{\beta,\alpha'}(x,x',\theta,\epsilon)\big]\notag\\
			\big[&g^{\text{R}}_{\beta',\alpha}(x',x,\theta',\epsilon-\omega)-g^{\text{A}}_{\beta',\alpha}(x',x,\theta',\epsilon-\omega)\big]\notag\\
			+\big[&f^{\text{R}}_{\beta,\beta'}(x,x',\theta,\epsilon)-f^{\text{A}}_{\beta,\beta'}(x,x',\theta,\epsilon)\big]\notag\\
			[&f^{\text{R}}_{\alpha',\alpha}(x',x,\theta',\epsilon-\omega)-f^{\text{A}}_{\alpha',\alpha}(x',x,\theta',\epsilon-\omega)\big]^{\dagger}\big\}\bigg],\label{eq:oc}
		\end{align}
		where we set $x'=x+0^+$ to evaluate the local value.
	}
	
	\showmatmethods{} % Display the Materials and Methods section
	
	\acknow{L.J.K. and J.L. acknowledge financial support from the Research Council of Norway through Grant No. 323766 and its Centres of Excellence funding scheme Project No. 262633 "QuSpin". L.J.K. acknowledge financial support from the Spanish Ministry for Science and Innovation—AEI Grant No. CEX2018-000805-M (through the “Maria de Maeztu” Programme for Units of Excellence in R\&D) and Grant No. RYC2021-031063-I funded by MCIN/AEI and “European Union Next Generation EU/PRTR”. B.L. acknowledges support from the National Natural Science Foundation of China (project 11904257). Y.T. acknowledges support from JSPS with Grants-in-Aid for Scientific Research \linak{(KAKENHI Grants No. 20H00131 and No. 23K17668, 24K00583)}. N.N. was supported by Japan Society for the JSPS KAKENHI Grant Numbers 24H00197 and 24H02231.
		N.N. was supported by the RIKEN Transformative Research Innovation Platform (TRIP) initiative.}
	
	\showacknow{} % Display the acknowledgments section

	\bibsplit[2]
	%Use \bibsplit to split the references from the body of the text. Value "[2]" represents the number of reference in the left column (Note: Please avoid single column figures & tables on this page.)
	
	% Bibliography

	\newpage
	\onecolumn
	
	\setcounter{equation}{0}
	\renewcommand{\theequation}{S.\arabic{equation}}
	\setcounter{figure}{0}
	\renewcommand{\thefigure}{S.\arabic{figure}}
	
	\section*{Supplemental Information}
	
	We here present: i) results for the local optical conductivity for additional values of the imaginary part of the energy $\epsilon+i\delta$ that demonstrate how the zero-energy zero-temperature local optical conductivity vanishes in the limit $\delta\to0$ in the case of $s$-wave or $d_{x^2-y^2}$-wave pairing and remains finite in the case of $d_{xy}$-wave pairing, ii) a further discussion of the experimental setup including constraints on the thickness of the superconductor and its contributions to the optical conductivity, iii) analytic expressions for the retarded McMillan Green's functions, and iv) the derivation of the local optical conductivity starting from the Kubo formula. 
	
	\subsection*{The optical conductivity in the limit of zero energy and zero temperature}
	
	\begin{figure*}[b!]
		\centering
		\includegraphics[width=\textwidth]{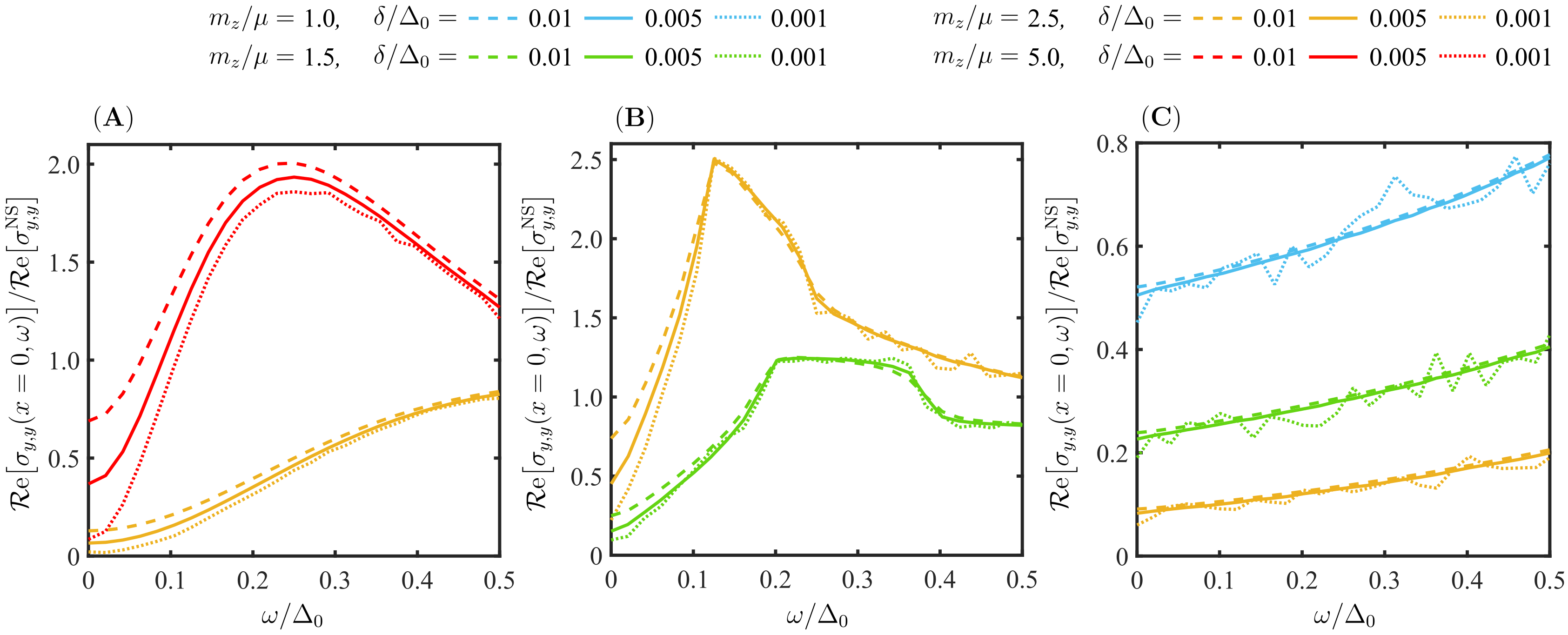}
		\caption{\textbf{Energy dependence of the local optical conductivity when decreasing $\delta$.} We consider the real part of the local optical conductivity $\mathcal{R}\text{e}[\sigma_{y,y}(x=0,\omega)]$ normalized by its normal-state value for $\Delta_0 = 0$ at the QAHI/TSC interface ($x=0$) and at zero temperature for $s$-wave [panel (\textbf{A})], $d_{x^2-y^2}$-wave [panel (\textbf{B})], and $d_{xy}$-wave [panel (\textbf{C)}] pairing for various values of the magnetization $m_z$ and imaginary part of the energy $\delta$. We consider $\Delta_0/\mu=10^{-3}\ll1$. For the $s$-wave and $d_{x^2-y^2}$-wave TSC, the local optical conductivity approaches zero in the limit $\omega\to0$ as $\delta$ decreases. For the $d_{x^2-y^2}$-wave TSC, the local optical conductivity remains finite.}
		\label{fig:SI01_v1}
	\end{figure*}
	
	In Fig.~\ref{fig:SI01_v1}, we demonstrate how the zero-temperature local optical conductivity close to $\omega=0$ depends on the value of the imaginary part of the energy $\epsilon+i\delta$. In the $s$-wave [panel~(\textbf{A}))] and $d_{x^2-y^2}$-wave [panel~(\textbf{B})] cases, the zero-temperature local optical conductivity decreases towards zero in the limit $\omega\to0$ when decreasing $\delta$. In these two cases, there are no nodal states at the small incidence angles $\theta$ that dominate the optical conductivity. Since the Majorana edge state can only contribute to the optical conductivity at finite energies \cite{He_PRL_2021}, the zero-energy zero-temperature optical conductivity therefore approaches zero in the limit $\delta\to0$. In the $d_{xy}$-wave case [panel~(\textbf{C})], the zero-energy zero-temperature local optical conductivity remains nearly unaffected by the decrease in $\delta$ due to the finite contribution from the nodal states at $\theta=0$. For all pairing symmetries, decreasing $\delta$ comes at the cost of noisy results. We therefore present data for the intermediate value of $\delta$ in the main text. 
	
	\subsection*{Further estimates regarding the experimental realization}
	
	As sketched in Fig.~\ref{fig:01}(\textbf{A}) in the main text, we consider a setup where a ferromagnetic insulator and a superconductor provides proximity-induced spin-splitting and superconductivity, respectively, to the topological insulator surface, and the MIM tip is scanned across the QAHI/TSC interface from above.
	As an upper limit for the superconductor thickness, we consider its skin depth. In the zero-temperature and zero-frequency limit, the skin depth equals the London penetration depth of the superconductor and can be estimated as
	\begin{align}
		\lambda=\frac{c}{\sqrt{\pi\Delta_0^{\text{SC}}\sigma_{\text{D}}/\epsilon_0\hbar}},
	\end{align}
	where $c$ is the speed of light, $\Delta_0^{\text{SC}}$ is the zero-temperature superconducting gap in the superconducting film, $\sigma_{\text{D}}$ is the Drude conductivity, $\epsilon_0$ is the electric permittivity, and $\hbar$ is the reduced Planck constant \cite{Hijano_PRB_2023}.
	
	An experiment found the proximity-induced gap from the high-$T_c$ $d$-wave superconductor Bi$_2$Sr$_2$CaCu$_2$O$_{8-\delta}$ onto the surface of the topological insulator Bi$_2$Se$_3$ to be $\Delta_0=13$~meV, while the reduced (bulk) superconducting gap in the superconductor was $\Delta_0^{\text{SC}}=27$~meV (45~meV) \cite{Zareapour_NatCommun_2012}.
	The Drude conductivity of Bi$_2$Sr$_2$CaCu$_2$O$_{8-\delta}$ at low temperatures is approximately 1500~$\Omega^{-1}\text{cm}^{-1}$ which corresponds to $\sigma_{\text{D}}/\epsilon_0=1.7\cdot10^{16}\:\text{s}^{-1}$ \cite{Liu_AnnPhys_2006}. Considering the reduced (bulk) gap, this corresponds to a skin depth $\lambda=200$~nm (160~nm) consistent with London penetration depths found for Bi$_2$Sr$_2$CaCu$_2$O$_{8-\delta}$ \cite{Tanner_Ferroelectrics_1996}.
	Since the skin depth only provides an upper limit, it is advantageous if the superconductor is as thin as possible. In experiments, $d$-wave superconductivity in Bi$_2$Sr$_2$CaCu$_2$O$_{8-\delta}$ has been measured down to the monolayer limit with an approximate thickness of 2~nm \cite{Yu_Nature_2019}. 
	
	Lee \textit{et al.} measured the optical conductivity with a resolution of 5~nm although the tip size was around 100~nm \cite{doi:10.1126/sciadv.abd1919}. In their measurement, the potential from the tip was reasonably uniform close to the 5~nm sized part of the tip where the signal is measured. It is thus reasonable to assume that the field traverses directly downwards over the spot size. We compare the spot size to the coherence length of proximity-induced superconductivity at the topological insulator surface, which can be estimated from $\xi_0=\hbar v_{\text{F}}/\pi\Delta_0$. We use the proximity-induced gap $\Delta_0=13$~meV from Bi$_2$Sr$_2$CaCu$_2$O$_{8-\delta}$ onto the surface of the topological insulator Bi$_2$Se$_3$ \cite{Zareapour_NatCommun_2012} and the Fermi velocity $v_{\text{F}}$ of the topological insulator surface.
	The Fermi velocity of Bi$_2$Te$_3$ was measured to be $v_{\text{F}}=4\cdot10^5$~m/s \cite{Qu_Science_2010} and takes a similar value for Bi$_2$Se$_3$ \cite{Sengupta_SemicondSciTech_2015}. The 5~nm spot size is therefore approximately 0.8$\xi_0$. 
	
	\begin{figure}[htb]
		\centering
		\includegraphics[width=\textwidth]{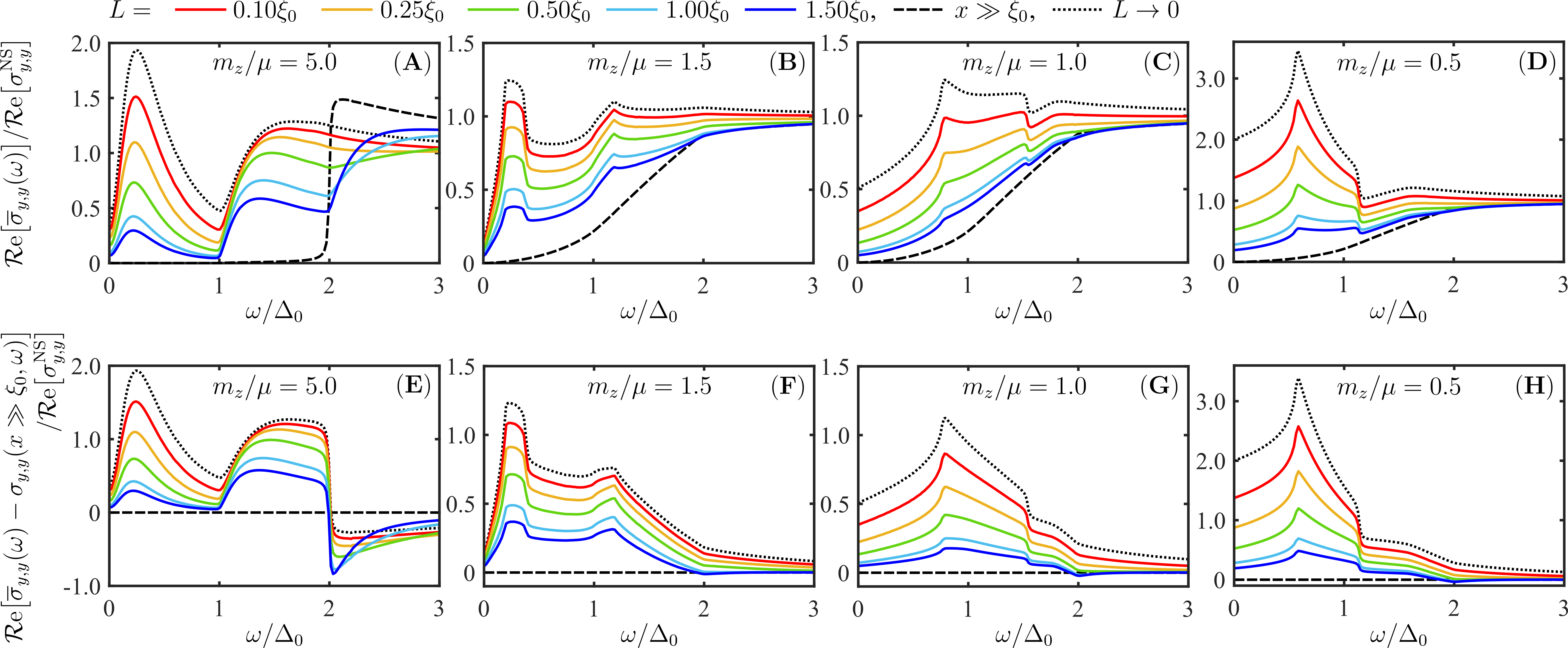}
		\caption{\textbf{Position-averaged optical conductivity.} We consider the real part of the averaged optical conductivity $\mathcal{R}\text{e}\big[\overline{\sigma}_{y,y}(\omega)\big]=\int_{0}^{\infty}dx\:\mathcal{R}\text{e}\big[\sigma_{y,y}(x,\omega)\big]\text{exp}[-(x/L)^2]/\int_{0}^{\infty}dx\:\text{exp}[-(x/L)^2]$ normalized by the normal-state value $\mathcal{R}\text{e}\big[\sigma_{y,y}^{\text{NS}}\big]$ in the (\textbf{A}) $s$-wave, (\textbf{B}) $d_{x^2-y^2}$-wave, and (\textbf{C})-(\textbf{D}) $d_{xy}$-wave case. The local optical conductivity at the interface ($x=0$) (black dotted line) and far away from the interface ($x\gg\xi_0$) (black dashed line) is plotted for comparison. Panel~(\textbf{E})-(\textbf{H}) shows the corresponding curves where the value far away from the interface $\mathcal{R}\text{e}\big[\sigma_{y,y}(x\gg\xi_0,\omega)\big]$ has been subtracted from results in panel (\textbf{A})-(\textbf{D}), respectively. We consider $\Delta_0/\mu=10^{-3}\ll1$ and $\delta/\Delta_0 = 5\cdot10^{-3}$.}
		\label{fig:SI02}
	\end{figure}
	
	To estimate how this spot size influences the signal from the Majorana edge mode, we have plotted the real part of the averaged optical conductivity  
	\begin{align}
		\mathcal{R}\text{e}\big[\overline{\sigma}_{y,y}(\omega)\big]=\frac{\int_{0}^{\infty}dx\:\mathcal{R}\text{e}\big[\sigma_{y,y}(x,\omega)\big]\text{exp}[-(x/L)^2]}{\int_{0}^{\infty}dx\:\text{exp}[-(x/L)^2]}
	\end{align}
	in Figs.~\ref{fig:SI02}(\textbf{A})-(\textbf{D}) for different values of $L$. We only include contributions from $x>0$, because the ferromagnetic insulator and topological insulator surface with proximity-induced magnetism are both insulating and can therefore not give contributions to the optical conductivity. To isolate how the optical conductivity is altered close to the QAHI/TSC interface, we subtract the optical conductivity far away from this interface ($x\gg\xi_0$) in panels~(\textbf{E})-(\textbf{H}). While the averaged optical conductivity is suppressed compared to the local value at the interface $(x=0)$, the peak in the optical conductivity originating from the Majorana edge mode remains considerable when $L$ is similar to the spot size, \lina{\textit{e.g.} for $L=\xi_0$ approximately 1/5 of the local value in panels (\textbf{E}), (\textbf{G}), and (\textbf{H}), and approximately 2/5 of the local value in panel (\textbf{F}). The exact ratio however depends on the pairing type, spot size, and value of $m_z/\mu$.} 
	
	So far, we only considered the contributions to the optical conductivity from the topological insulator surface. We now estimate how this conductivity compares to the optical conductivity of the $d$-wave superconductor deposited on top of the TI. In the low-frequency limit, the optical conductivity of $d$-wave superconductors have been found to be of the order of $1000\:\Omega^{-1}$cm$^{-1}$ \cite{Puchkov_PRL_1996,Liu_JPhysCondensMatter_1999} in the presence of impurities \cite{Dora_CurrApplPhys_2006}. \lina{To find the conductance of the thin-film superconductor, we therefore have to multiply the three-dimensional conductivity with the thickness of the superconductor}, $nc$, where $n$ is the number of layers and $c$ is the out-of-plane lattice constant here taken to be $30$~\AA.
	The low-energy conductance of Bi$_2$Se$_3$ thin films was measured to be approximately 0.3$~\Omega^{-1}$ and nearly independent of the thickness since the dominant contribution comes from the surface state \cite{ValdesAguilar_PRL_2012}.
	The conductance of \lina{a stack of} $n$ superconducting layers normalized by the conductance of the topological insulator surface in the normal state is thus $G^{\text{SC}}_{y,y}/G^{\text{NS}}_{y,y}=n\cdot30\:\text{\AA}\cdot1000\:\Omega^{-1}\text{cm}^{-1}/0.3\:\Omega^{-1}=n\cdot 10^{-3}$. 
	
\begin{figure} [htb]
		\centering
		\includegraphics[width=\textwidth]{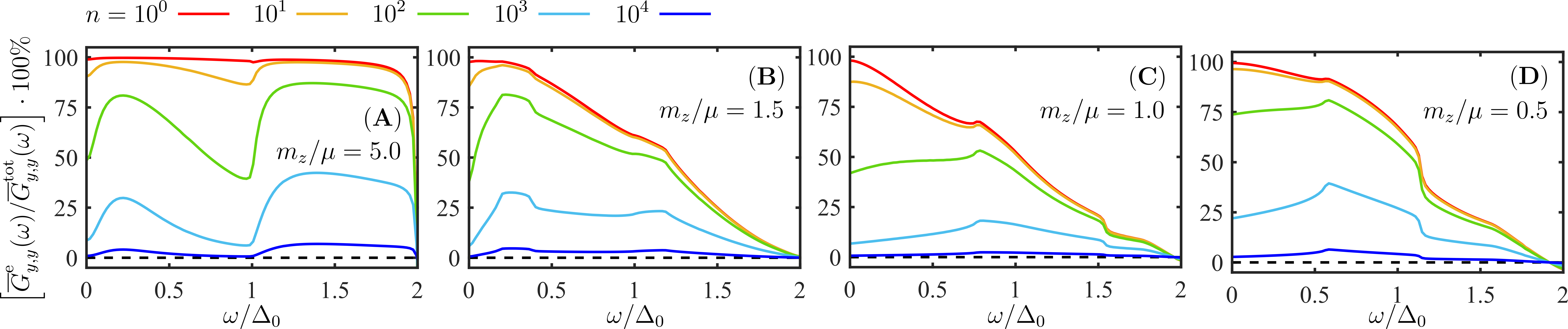}
		\caption{\lina{\textbf{Percentage contribution from the edge mode.} We consider the conductance stemming from the edge mode $\overline{G}^{\text{e}}_{y,y}(\omega)$ as a percentage of the total conductance $\overline{G}^{\text{tot}}_{y,y}(\omega)$ from the \lina{thin-film} superconductor and topological insulator surface in the $s$-wave [panel (\textbf{A})], $d_{x^2-y^2}$-wave [panel (\textbf{B})], and $d_{xy}$-wave [panels (\textbf{C})-(\textbf{D})] case. The conductance of the edge mode relates to the optical conductivity in Fig.~\ref{fig:SI02}(\textbf{E})-(\textbf{H}), respectively, for a realistic spot size $L=\xi_0$. We consider various values for the number $n$ of atomic layers in the superconductor, $G^{\text{SC}}_{y,y}/G_{y,y}^{\text{NS}}=n\cdot10^{-3}$, $\Delta_0/\mu=10^{-3}\ll1$ and $\delta/\Delta_0 = 5\cdot10^{-3}$.}}
		\label{fig:SI03}
	\end{figure}

	In the main text and in Fig.~\ref{fig:SI02}, we have considered the optical conductivity of the two-dimensional topological insulator surface with proximity-induced superconductivity normalized by the optical conductivity of the same surface in the normal-state. We assume that the thickness $t$, representing the thickness over which the surface states extends along the $z$ axis, is approximately equal in the superconducting and normal state. In this case, the conductance ratio $G_{y,y}(x,\omega)/G_{y,y}^{\text{NS}}=\mathcal{R}\text{e}\big[\sigma_{y,y}(x,\omega)\big]/\mathcal{R}\text{e}\big[\sigma^{\text{NS}}_{y,y}\big]$ and its position averaged counterpart $\overline{G}_{y,y}(\omega)/G_{y,y}^{\text{NS}}=\mathcal{R}\text{e}\big[\overline{\sigma}_{y,y}(\omega)\big]/\mathcal{R}\text{e}\big[\sigma^{\text{NS}}_{y,y}\big]$ equal the ratios of the corresponding optical conductivities due to the thicknesses cancelling out. 
	This allows us to estimate the contribution from the edge mode $\overline{G}^{\text{e}}_{y,y}(\omega)=\overline{G}_{y,y}(\omega)-G_{y,y}(x\gg\xi_0,\omega)$ as a percentage of the total conductance $\overline{G}_{y,y}^{\text{tot}}(\omega)=\overline{G}_{y,y}(\omega)+G_{y,y}^{\text{SC}}$ in Fig.~\ref{fig:SI03} for various values of $n$. We consider a realistic spot size where $L=\xi_0$. The conductance peak resulting from the edge states give the dominating contribution to the total signal for a superconducting film with $n$ up to the order of hundred atomic layers \lina{$n\sim\mathcal{O}(10^2)$}. For thicker superconductor films, \textit{e.g.} of the order of hundred micrometers or \lina{$n\sim\mathcal{O}(10^4)$} atomic layers considered in Ref.~\cite{Zareapour_NatCommun_2012}, the peak signal from the edge mode is still a few percent of the total signal and should therefore be possible to resolve in experiments. \lina{In the $d$-wave case, the contribution from the edge mode is percentage-wise larger compared to the total signal for small values of $\omega$ due to contributions from the nodes.}
	
	\subsection*{Analytic expressions for the retarded McMillan Green's functions}
	
	The $2\times2$ retarded McMillan Green's function matrices in Eq.~(25) in the main text are given by
	\begin{align}
		g^{\text{R}}(x_1,x_2,\theta,\epsilon&+i\delta) = \notag\\
		-\frac{i}{2A\cos(\theta)}
		\bigg[&\left(\frac{1}{1-\Gamma_{\horpm}^2}e^{ik_x^{\text{SCe}\horpm}|x_1-x_2|}+\frac{\Gamma_{\horpm} C_{\hormp}}{1-\Gamma_{\horpm}^2}e^{ik_x^{\text{SCe}\horpm}x_{1(2)}-ik_x^{\text{SCh}\horpm}x_{2(1)}}\right)
		\begin{pmatrix}
			1 & \horpm e^{\hormp i\theta}\\
			\horpm e^{\horpm i\theta} & 1
		\end{pmatrix}\notag\\
		+&\left(\frac{\Gamma_{\hormp}^2}{1-\Gamma_{\hormp}^2}e^{- ik_x^{\text{SCh}\hormp}|x_1-x_2|}+\frac{\Gamma_{\hormp} C_{\horpm}}{1-\Gamma_{\hormp}^2}e^{-ik_x^{\text{SCh}\hormp}x_{1(2)}+ik_x^{\text{SCe}\hormp}x_{2(1)}}\right)
		\begin{pmatrix}
			1 & \hormp e^{\horpm i\theta}\\
			\hormp e^{\hormp i\theta} & 1
		\end{pmatrix}
		\bigg],\:x_1 \horgl x_2\\
		f^{\text{R}}(x_1,x_2,\theta,\epsilon&+i\delta) = \notag\\
		-\frac{i}{2A\cos(\theta)}\bigg[&\left(\frac{\Gamma_{\horpm}}{1-\Gamma_{\horpm}^2}e^{ik_x^{\text{SCe}\horpm}|x_1-x_2|}+1(\Gamma_-^2)\frac{C_{\hormp}}{1-\Gamma_{\horpm}^2}e^{ik_x^{\text{SCe}\horpm}x_{1(2)}-ik_x^{\text{SCh}\horpm}x_{2(1)}}\right)
		\begin{pmatrix}
			\hormp e^{\hormp i\theta} & 1\\
			-1 & \horpm e^{\horpm i\theta}
		\end{pmatrix}\notag\\
		+&\left(\frac{\Gamma_{\hormp}}{1-\Gamma_{\hormp}^2}e^{- ik_x^{\text{SCh}\hormp}|x_1-x_2|}+\Gamma_{-}^2(1)\frac{ C_{\horpm}}{1-\Gamma_{\hormp}^2}e^{-ik_x^{\text{SCh}\hormp}x_{1(2)}+ik_x^{\text{SCe}\hormp}x_{2(1)}}\right)
		\begin{pmatrix}
			\horpm e^{\horpm i\theta} & 1\\
			-1 & \hormp e^{\hormp i\theta}
		\end{pmatrix}
		\bigg],\:x_1\horgl x_2,\\
		\underline{g^{\text{R}}}(x_1,x_2,\theta,\epsilon&+i\delta) =\notag\\
		-\frac{i}{2A\cos(\theta)}\bigg[&\left(\frac{\Gamma_{\horpm}^2}{1-\Gamma_{\horpm}^2}e^{ ik_x^{\text{SCe}\horpm}|x_1-x_2|}+\frac{\Gamma_{\horpm} C_{\hormp}}{1-\Gamma_{\horpm}^2}e^{ik_x^{\text{SCe}\horpm}x_{1(2)}-ik_x^{\text{SCh}\horpm}x_{2(1)}}\right)
		\begin{pmatrix}
			1 & \hormp e^{\horpm i\theta}\\
			\hormp e^{\hormp i\theta} & 1
		\end{pmatrix}\notag\\
		+&\left(\frac{1}{1-\Gamma_{\hormp}^2}e^{- ik_x^{\text{SCh}\hormp}|x_1-x_2|}+\frac{\Gamma_{\hormp} C_{\horpm}}{1-\Gamma_{\hormp}^2}e^{-ik_x^{\text{SCh}\hormp}x_{1(2)}+ik_x^{\text{SCe}\hormp}x_{2(1)}}\right)
		\begin{pmatrix}
			1 & \horpm e^{\hormp i\theta}\\
			\horpm e^{\horpm i\theta} & 1
		\end{pmatrix}
		\bigg],\:x_1\horgl x_2\\
		\underline{f^{\text{R}}}(x_1,x_2,\theta,\epsilon&+i\delta) = \notag\\
		-\frac{i}{2A\cos(\theta)}\bigg[&\left(\frac{\Gamma_{\horpm}}{1-\Gamma_{\horpm}^2}e^{ ik_x^{\text{SCe}\horpm}|x_1-x_2|}+\Gamma_{+}^{2}(1)\frac{ C_{\hormp}}{1-\Gamma_{\horpm}^2}e^{ik_x^{\text{SCe}\horpm}x_{1(2)}-ik_x^{\text{SCh}\horpm}x_{2(1)}}\right)
		\begin{pmatrix}
			\hormp e^{\horpm i\theta} & -1\\
			1 & \horpm e^{\hormp i\theta}
		\end{pmatrix}\notag\\
		+&\left(\frac{\Gamma_{\hormp}}{1-\Gamma_{\hormp}^2}e^{- ik_x^{\text{SCh}\hormp}|x_1-x_2|}+1(\Gamma_+^2)\frac{C_{\horpm}}{1-\Gamma_{\hormp}^2}e^{-ik_x^{\text{SCh}\hormp}x_{1(2)}+ik_x^{\text{SCe}\hormp}x_{2(1)}}\right)
		\begin{pmatrix}
			\horpm e^{\hormp i\theta} & -1\\
			1 & \hormp e^{\horpm i\theta}
		\end{pmatrix}\bigg],\:x_1\horgl x_2.
	\end{align}
	These are valid inside the superconducting region [$(x_1,x_2)>0$]. The coefficients are given by
	\begin{align}
		C_{\horpm} &=-\frac{(1-\sigma_{\text{NS}})e^{i\eta}\Gamma_{\horpm}+\Gamma_{\hormp}}{1+(1-\sigma_{\text{NS}})e^{i\eta}\Gamma_+\Gamma_-}.
	\end{align}
	where 
	\begin{align}
		\sigma_{\text{NS}} = \frac{\cos^2(\theta)}{\cosh^2(\kappa^{\text{FI}}_x d)\cos^2(\theta)+\left(\frac{1-\gamma^2}{1+\gamma^2}\right)^2\sinh^2(\kappa^{\text{FI}}_x d)\sin^2(\theta)}.
	\end{align}
	is the normal-state conductance through the ferromagnetic region \cite{Tanaka_PRL_2009,Linder_PRL_2010}, and $\eta$ is defined by
	\begin{align}
		e^{i\eta} = \frac{m_z \cos(\theta)+i\mu\sin(\theta)}{m_z \cos(\theta)-i\mu\sin(\theta)},
	\end{align}
	where $m_z$ is the magnetization in the ferromagnetic region ($-d<x<0$) and $\mu$ is the chemical potential in the non-magnetic non-superconducting region ($x<-d$) and in the superconducting region ($x>0$). 
	
	\subsection*{The Kubo formula for the optical conductivity}
	
	To evaluate the local optical conductivity, we start from the Kubo formula \cite{Mahan_Book_2000}
	\begin{align}
		&\sigma_{i,j}(x,x',q_y,i\omega_n) =
		-\frac{1}{\omega_n L_y}\int_0^{\beta} d\tau\: e^{i\omega_n \tau}\left<T_{\tau}j_{i}(x,-q_y,\tau)j_{j}(x',q_y,0)\right>
	\end{align}
	for a system of length $L_y$ with periodic boundary conditions along $\ve{y}$. Above, $x$ and $x'$ are real space coordinates, $q_y$ is the momentum along $\ve{y}$, $\omega_n$ is the Matsubara frequency, $\beta=1/k_{\text{B}}T$, and $T_{\tau}$ is the time ordering operator for the imaginary time~$\tau$. To obtain an expression for the real part of the optical conductivity in terms of the retarded and advanced Green's functions, we proceed as follows: First, we insert the inverse Fourier transform
	\begin{align}
		j_{i}(x,q_y,\tau)=\int dy\: e^{iq_y y}j_{i}(\ve{r},\tau)
	\end{align}
	of the current operator, where 
	\begin{align}
		j_i(\ve{r},\tau)&=-eA\sum_{\alpha,\beta}\psi_{\alpha}^{\dagger}(\ve{r},\tau)(\sigma_i)_{\alpha,\beta}\psi_{\beta}(\ve{r},\tau)
	\end{align}
	and $\ve{r}=(x,y)$.
	Next, using Wick's theorem, we write the expectation values over the four fermion operators in terms of the Matsubara Green's functions
	\begin{align}
		\mathcal{G}_{\alpha,\beta}(\ve{r},\ve{r}',\tau-\tau')& = -\left<T_{\tau}\psi_{\alpha}(\ve{r},\tau)\psi_{\beta}^{\dagger}(\ve{r}',\tau')\right>,\\
		\mathcal{F}_{\alpha,\beta}(\ve{r},\ve{r}',\tau-\tau')& = -\left<T_{\tau}\psi_{\alpha}(\ve{r},\tau)\psi_{\beta}(\ve{r}',\tau')\right>.
	\end{align}
	We introduce the center of mass coordinate $Y=(y+y')/2$ and the relative coordinate $y_{\text{rel}}=y-y'$, and assume that the Matsubara Green's functions are independent of $Y$. We insert the Fourier transform
	\begin{align}
		\mathcal{G}_{\alpha,\beta}(\ve{r},\ve{r}',\tau) &= \frac{1}{\beta}\sum_n e^{-i\omega_n\tau}\mathcal{G}_{\alpha,\beta}(\ve{r},\ve{r}',i\omega_n),
	\end{align}
	of the Matsubara Green's functions, and use the relation
	\begin{align}
		\frac{1}{\beta}\int_0^{\beta}d\tau\:e^{i(\omega_n+p_m-p_{m'})\tau}=\delta_{p_{m'},\omega_n+p_m}
	\end{align}
	to replace the imaginary time integral with sums over Matsubara frequencies.
	Assuming that $x\neq x'$, we eliminate terms where neither of the Matsubara Green's functions depend on $i\omega_n$. We then insert
	\begin{align}
		&\mathcal{G}(x,x',y_{\text{rel}},i\omega_n)
		=
		-\frac{1}{2i\pi}\int_{-\infty}^{\infty}d\epsilon\:\frac{[g^{\text{R}}(x,x',y_{\text{rel}},\epsilon)-g^{\text{A}}(x,x',y_{\text{rel}},\epsilon)]}{i\omega_n-\epsilon}
	\end{align}
	[and similarly for $\mathcal{F}(x,x',y_{\text{rel}},i\omega_n)$], and evaluate the Matsubara sum
	\begin{align}
		-\frac{1}{\beta}\sum_m \frac{1}{(ip_m -\epsilon)(i\omega_n+ip_m-\epsilon')} = \frac{f_{\text{FD}}(\epsilon')-f_{\text{FD}}(\epsilon)}{i\omega_n-\epsilon'+\epsilon},
	\end{align}
	via contour integration. Above, $f_{\text{FD}}(\epsilon)$ is the Fermi-Dirac distribution. This allows us to perform the analytic continuation $i\omega_n\to\omega+i\delta$. We insert the Fourier transform 
	\begin{align}
		g^{\text{R}}(x,x',y_{\text{rel}},\epsilon)=\int \frac{dk_y}{2\pi}\: e^{-ik_y y}g^{\text{R}}(x,x',k_y,\epsilon)
	\end{align}
	[and similarly for $f^{\text{R}}(x,x',y_{\text{rel}},\epsilon)$], and use the relation
	\begin{align}
		\int dy_{\text{rel}}\:e^{i(k_y-k'_y)y}=2\pi\delta_{k_y,k'_y}. 
	\end{align}
	Finally, we introduce a cutoff $\pm\mu/A$, so that the integrals over the momenta $k_y$ can be written as an integral over the incidence angles,
	\begin{align}
		\int\frac{dk_y}{2\pi}\:\to\frac{\mu}{2A}\int_{-\pi/2}^{\pi/2}\frac{d\theta}{\pi}\:\cos(\theta).
	\end{align}
	We also average over $q_y$ by introducing a second integral $\int_{-\pi/2}^{\pi/2}(d\theta'/\pi)\cos(\theta')$.
	Since we are interested in the real part of the optical conductivity, the relation
	\begin{align}
		&\Real\left[\lim_{\delta\to0^+}\int_{-\infty}^{\infty}d\epsilon\:\frac{F(\epsilon)}{\epsilon+i\delta-(\epsilon'-\omega)}\right]
		=\Real[-i\pi F(\epsilon'-\omega)],
	\end{align}
	greatly simplifies the expression.
	The resulting real part of the local optical conductivity is given in Eq.~(31) in the main text.

\end{document}